\documentclass[twocolumn]{aastex631}

\usepackage{amsmath}
\usepackage{natbib}
\usepackage{xcolor}
\newcommand{\lcdm}{$\Lambda$CDM}

\newcommand{\om}{\Omega_{m0}}

\newcommand{\FT}[1]{}

\newcommand{\Mgii}{Mg\,\textsc{ii}}
\newcommand{\Civ}{C\,\textsc{iv}}

\begin{document}

\title{Effect of extinction on quasar luminosity distances determined from UV and X-ray flux measurements}

\correspondingauthor{Michal Zaja\v{c}ek}
\email{zajacek@physics.muni.cz}

\author[0000-0001-6450-1187]{Michal Zaja\v{c}ek}
\affiliation{Department of Theoretical Physics and Astrophysics, Faculty of Science, Masaryk University, Kotl\'a\v{r}sk\'a 2, 611 37 Brno, Czech Republic}

\author[0000-0001-5848-4333]{Bo\.{z}ena Czerny}
\affiliation{Center for Theoretical Physics, Polish Academy of Sciences, Al.\ Lotnik\'{o}w 32/46, 02-668 Warsaw, Poland}

\author[0000-0001-5512-2716]{Narayan Khadka}
\affiliation{Department of Physics, Bellarmine University, 2001 Newburg Rd, Louisville, KY 40205, USA}
\affiliation{Department of Physics and Astronomy, Stony Brook University, Stony Brook, NY 11794, USA}

\author[0000-0002-7843-7689]{Mary Loli Mart\'inez-Aldama}
\affiliation{Astronomy Department, Universidad de Concepción, Casilla 160-C, Concepción 4030000, Chile}
\affiliation{Instituto de Física y Astronomía, Facultad de Ciencias, Universidad de Valparaíso, Gran Bretaña 1111, Playa Ancha, Valparaíso, Chile}

\author[0000-0002-1173-7310]{Raj Prince}
\affiliation{Center for Theoretical Physics, Polish Academy of Sciences, Al.\ Lotnik\'{o}w 32/46, 02-668 Warsaw, Poland}

\author[0000-0002-5854-7426]{Swayamtrupta Panda}\thanks{CNPq Fellow}
\affiliation{Laborat\'orio Nacional de Astrof\'isica - MCTI, R. dos Estados Unidos, 154 - Na\c{c}\~oes, Itajub\'a - MG, 37504-364, Brazil}

\author[0000-0002-7307-0726]{Bharat Ratra}
\affiliation{Department of Physics, Kansas State University, 116 Cardwell Hall, Manhattan, KS 66506, USA}



\begin{abstract}

In Khadka et al. (2023), a sample of X-ray-detected reverberation-mapped quasars was presented and applied for the comparison of cosmological constraints inferred using two well-established relations in AGN -- the X-ray/UV luminosity ($L_{X}-L_{UV}$) relation and the broad-line region radius-luminosity ($R-L$) relation. $L_{X}-L_{UV}$ and $R-L$ luminosity distances to the same quasars exhibit a distribution of their differences that is generally asymmetric and positively shifted for the six cosmological models we consider. We demonstrate that this behaviour can be interpreted qualitatively to arise as a result of the dust extinction of UV/X-ray quasar emission. We show that the extinction always contributes to the non-zero difference between $L_{X}-L_{UV}$-based and $R-L$-based luminosity distances and we derive a linear relationship between the X-ray/UV colour index $E_{X-UV}$ and the luminosity-distance difference, which also depends on the value of the $L_{X}-L_{UV}$ relation slope. Taking into account the median and the peak values of the luminosity-distance difference distributions, the average X-ray/UV colour index falls in the range of $\overline{E}_{X-UV}=0.03-0.28$ mag for the current sample of 58 sources. This amount of extinction is typical for the majority of quasars and it can be attributed to the circumnuclear and interstellar media of host galaxies. After applying the standard hard X-ray and far-UV extinction cuts, heavily extincted sources are removed but overall the shift towards positive values persists. The effect of extinction on luminosity distances is more pronounced for the $L_{X}-L_{UV}$ relation since the extinction of UV and X-ray emissions both contribute.

\end{abstract}


\section{Introduction}

The spatially-flat $\Lambda$CDM cosmological model \citep{Peebles1984} is largely consistent with most lower-redshift, $z\lesssim 2.3$, observations \citep{Yuetal2018, eBOSSCollaboration2021, Broutetal2022} as well as with high-redshift cosmic microwave background (CMB) data at $z\sim 1100$ \citep{PlanckCollaboration2020}. However, there are several potential tensions between flat $\Lambda$CDM parameter values inferred using different techniques \citep{PerivolaropoulosSkara2021, Morescoetal2022, Abdallaetal2022, HuWang2023}. These can be addressed by improving the accuracy and precision of established cosmological probe measurements, and also by looking for alternative cosmological probes, especially in the redshift range between nearby data and CMB data. 

Active galactic nuclei, especially bright quasars \citep[QSOs;][]{2021bhns.confE...1K,2023arXiv230615082Z}, appear to be promising alternative probes due to their broad redshift coverage, ranging from the nearby Universe \citep[$z=0.00106$ for NGC4395;][]{2019MNRAS.486..691B} to $z\approx 7.642$ \citep[J0313–1806;][]{2021ApJ...907L...1W}. For cosmological applications, so far three types of QSO data have been more widely utilized: (i) QSO angular size observations \citep{Caoetal2017, Ryanetal2019, Caoetal2020, Caoetal2021a,Caoetal_2021b, Caoetal_2021c,Lianetal2021}; (ii) data based on the non-linear relation between QSO X-ray and UV luminosities, the $L_{X}-L_{UV}$ relation \citep{RisalitiLusso2015, RisalitiLusso2019, KhadkaRatra2020a, KhadkaRatra2020b, KhadkaRatra2021a, KhadkaRatra2022, Lussoetal2020, Lietal2021, Rezaeietal2022, HuWang2022, Colgainetal2022, dainotti2022, Petrosianetal2022, Khadkaetal2023}; and, (iii) data based on the correlation between the rest-frame broad-line region (BLR) time delay and the monochromatic luminosity, the $R-L$ relation \citep{2019ApJ...883..170M,2019FrASS...6...75P, panda2019, Michal2021, khadka2021, 2022MNRAS.513.1985K, Khadkaetal2022, Czerny2021, Czernyetal2022, Panda_2022FrASS...950409P, CaoRatra2022, CaoRatra2023, Caoetal2022,Caoetal2023, Panda_Marziani_2023}. In addition to these methods, \citet{2002ApJ...581L..67E} suggested using angular diameters of the BLR to measure the cosmological constant $\Lambda$ \citep[also see][]{2020NatAs...4..517W} and \citet{1999MNRAS.302L..24C} as well as \citet{2007MNRAS.380..669C} discussed application of continuum reverberation mapping to measure the Hubble constant. Also, the H0LiCOW (H0 Lenses in COSMOGRAIL's Wellspring) program based on time delays between lensed images of QSOs has measured the Hubble constant \citep{Birreretal2020}.

QSO standardization based on the $L_{X}-L_{UV}$ relation and the constructed Hubble diagram has led to claims of strong cosmological constraints and tension with the $\Lambda$CDM model with non-relativistic matter density parameter $\om \sim 0.3$ \citep{RisalitiLusso2019, Lussoetal2020}. However, the analyses of \citet{RisalitiLusso2019} and \citet{Lussoetal2020} were approximate and based on incorrect assumptions \citep{KhadkaRatra2020a, KhadkaRatra2020b, KhadkaRatra2021a, KhadkaRatra2022, Banerjeeetal2021, Petrosianetal2022}, i.e. cosmological parameters and $L_{X}-L_{UV}$ relation parameters were constrained within the non-flat $\Lambda$CDM model, hence the results were model-dependent. The correct technique for analysis of these data was developed by \citet{KhadkaRatra2020a} and here we outline it as follows: given the current quality of these data, one must use them to simultaneously determine the $L_X-L_{UV}$ relation parameters and the cosmological model parameters, and one must also study a number of different cosmological models to determine whether $L_X-L_{UV}$ relation parameter values are independent of the assumed cosmological model. If the $L_X-L_{UV}$ relation parameter values are independent of the assumed cosmological model, the QSOs are standardizable and the circularity problem is circumvented. We emphasize, however, that when correctly analyzed, the most recent \citet{Lussoetal2020} data compilation is not standardizable \citep{KhadkaRatra2021a, KhadkaRatra2022}, because the $L_X-L_{UV}$ relation parameters depend on the assumed cosmological model as well as on redshift \citep{KhadkaRatra2021a, KhadkaRatra2022}. \citet{KhadkaRatra2022} discovered that the largest of the seven sub-samples in the \citet{Lussoetal2020} QSO compilation, the SDSS-4XMM one that contains about 2/3 of the total QSOs, has an $L_{X}-L_{UV}$ relation that depends on the cosmological model and on redshift and is the main source of the problem with the \citet{Lussoetal2020} data.

On the other hand, the BLR $R-L$ relation parameters generally appear independent of the adopted cosmological model \citep{khadka2021, 2022MNRAS.513.1985K, Khadkaetal2022, Caoetal2022}. Cosmological constraints are weak, but for \Mgii\ (at 2798\,\AA\, in the rest frame) and \Civ\ QSOs (at 1549\,\AA\, in the rest frame) they are consistent with those from better-established probes \citep{Khadkaetal2022, Caoetal2022}. However, there is a 2$\sigma$ tension between the constraints given by lower-redshift H$\beta$ QSOs (at 4861\,\AA\, in the rest frame) and those from the better-established probes \citep{2022MNRAS.513.1985K}. The H$\beta$ QSO sample yields weak cosmological constraints with a preference for decelerated cosmological expansion. This tension, and possible systematic issues, related to H$\beta$ QSOs will need to be addressed when more reverberation-mapped (RM) QSOs are available, e.g., from the upcoming Vera C.\ Rubin telescope and its Legacy Survey of Space and Time, LSST \citep[see e.g.][]{2019ApJ...873..111I,2019FrASS...6...75P,2023arXiv230108975C} and the Sloan Digital Sky Survey V (SDSS-V) Black Hole Mapper \citep{almeida2023}. 

In summary, both $L_{X}-L_{UV}$ relation and $R-L$ relation data provide cosmological constraints, which is encouraging, but they have their systematic problems that require further study. To uncover the systematic issues we looked for a sample of X-ray-detected quasars that have been reverberation-mapped in the UV domain using the \Mgii\ line. For a such sample, in principle, both $L_{X}-L_{UV}$ and $R-L$ relations should be applicable. To this goal, we provided a sample of 58 X-ray detected RM \Mgii\ QSOs in \citet{Khadkaetal2023}, from which we measured both $L_{X}-L_{UV}$ relation and $R-L$ relation parameter values that are consistent with values measured using larger samples. The sample was quite limited but we stress that for each of the sources we could determine the luminosity distance from these two methods independently. The goal was to compare cosmological constraints inferred from the QSOs for both of these relations not just in the overall statistical study of two independent samples but for the same sample. Using six cosmological models, we found that both $R-L$ and $L_{X}-L_{UV}$ relations are standardizable. The main result of this study was that, while the $R-L$ relation measurements favoured a smaller $\om$ value consistent with the $\om \sim 0.3$ value measured using better-established cosmological probes, the $L_{X}-L_{UV}$ relation measurements favoured a value of $\om$ larger than that inferred from the better-established cosmological probes.

In \citet{Khadkaetal2023} we showed that this is in agreement with the tendency for luminosity distances based on $L_{X}-L_{UV}$ relation data to have a mean value smaller than luminosity distances based on $R-L$ relation data for the same QSOs. The median $L_{X}-L_{UV}$ relation luminosity distance values, on the other hand, are mostly larger except for the flat and non-flat XCDM models. We note that these results are based on the distribution of the luminosity-distance difference $(\log{D_{L,R-L}}-\log{D_{L,L_{X}-L_{UV}}})$ normalized by the square root of the quadratic sum of the corresponding uncertainties.

Here we revisit the non-zero median (mean) of the difference of the luminosity distances based on $L_{X}-L_{UV}$ relation and $R-L$ relation data sets. It is expected that, in the absence of systematic effects, the median (mean) of the distance difference distribution should vanish within uncertainties since we construct the distribution from the same sample of sources and the luminosity distance to a particular source must be the same for both methods. The offset from zero hints at a systematic effect in one or both data sets and demands further analysis.

We propose that the offset in the $\Delta \log{D_{\rm L}}\equiv \log{D_{L,L_{X}-L_{UV}}}-\log{D_{L,R-L}}$ distribution may be attributed to the extinction (absorption and scattering) of UV and X-ray photons along the line of sight. Since the extinction curve generally increases from the UV (2500\,\AA) to the soft X-ray bands (2 keV), we evaluate the differential extinction or the optical depth difference between X-ray and UV bands, $\tau_{X}-\tau_{UV}$. Specifically, we derive a simple analytical relation between $\tau_{X}-\tau_{UV}$ and the slope $\gamma'$ of the $L_{X}-L_{UV}$ relation as well as the luminosity-distance difference $\Delta \log{D_{\rm L}}$, i.e. $\tau_{X}-\tau_{UV}\simeq 4.6(\gamma'-1)\Delta \log{D_{\rm L}}$. This way the mean differential extinction effect, or the X-ray/UV colour index, can be directly inferred for the current sample of 58 X-ray RM QSOs as well as any future sample of such sources.

We show that the asymmetry and the positive shift of $\Delta \log{D_{\rm L}}$ distributions for all the considered cosmological models, including the flat $\Lambda$CDM model, with prevailing positive medians, can be explained by a mild extinction of the UV and X-ray flux densities from QSOs. From all the median and the peak values of the $\Delta \log{D_{\rm L}}$ distributions, the X-ray/UV colour index falls in the range of $\sim 0.03-0.28$ mag for the sample of 58 X-ray detected, RM sources \citep{Khadkaetal2023}. We demonstrate that this amount of extinction is typical for the majority of QSOs, originates in the circumnuclear and interstellar media of host galaxies \citep{2004MNRAS.348L..54C}, and is only slightly alleviated by standard hard X-ray and far-UV extinction cuts \citep{Lussoetal2020}. Since the $L_{X}-L_{UV}$ relation employs both UV and X-ray flux density measurements, it turns out to be more affected by extinction, which can address some of the previously mentioned problems with its standardization.

Our paper is structured as follows. In Section~\ref{sec:data} we describe the characteristics of the dataset of X-ray RM QSOs. In Section~\ref{sec:models} we summarize the calculation method for luminosity distances and the different cosmological models we utilize. Subsequently, in Section~\ref{sec:methods}, we derive the relation between the X-ray and UV optical depth difference (colour index) and the luminosity-distance difference. Our main results are presented in Section~\ref{sec:results}. We provide additional discussion of the source selection and our results in Section~\ref{sec:discussion} and conclude in Section~\ref{sec:conclusions}.

The reader may also find additional derivations of luminosity distances including the extinction law in Appendix~\ref{appendix: formulae}, whereas the distributions related to the luminosity-distance difference formula are presented in Appendix~\ref{appendix: delta-flux}. Additional distributions of the X-ray/UV colour index for different cosmological models are shown in Appendix~\ref{appendix: E_X_UV} and the detailed investigation of reddening using SDSS magnitudes is presented in Appendix~\ref{appendix:reddening}.  

\begin{figure*}
    \includegraphics[width=\textwidth]{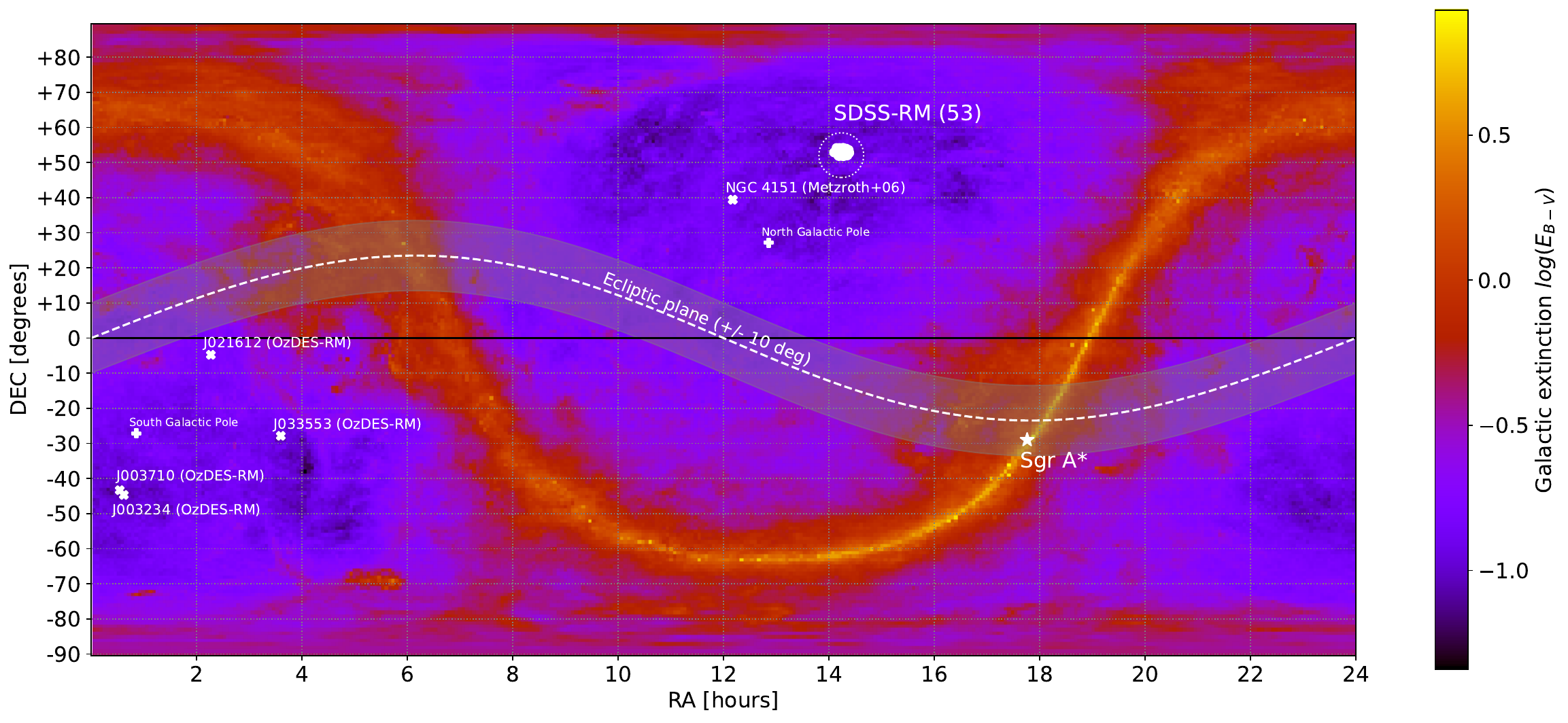}
    \caption{The position of the sample of 58 X-ray detected RM sources on the sky. Right ascension (RA expressed in hours) is along the $x$-axis, while declination (DEC expressed in degrees) is along the $y$-axis. The white dotted circle in the North hemisphere denotes the sample of 53 SDSS-RM sources \citep{2016ApJ...818...30S, shen_2019, Homayouni2020}, while the five remaining sources \citep{2006ApJ...647..901M,Zhefu2021} are more scattered to the south. The figure also shows the dust extinction map of the Milky Way (red-orange band) colour-coded using the logarithm of the $(B-V)$ colour index $E_{B-V}$ according to \citet{1998ApJ...500..525S}. The position of the Galactic centre (Sgr~A*) is marked by a white star, while the North and the South Galactic poles are depicted by plus signs. The Ecliptic plane and the surrounding band of $\pm$ 10 degrees are represented by the white dashed line and the grey shaded area, respectively.}
    \label{fig_quasars_dustmap}
\end{figure*}

\section{Data description}
\label{sec:data}

Our sample consists of 58 sources that were both (i) reverberation mapped using UV continuum and \Mgii\ emission-line light curves, and (ii) X-ray detected at 2 keV. In this regard, the sample is complete and due to its small size, we are not imposing further selection criteria unless stated otherwise. There are altogether 59 measurements of BLR time delays since the time delay for NGC 4151 was measured twice \citep{2006ApJ...647..901M}. The sample is described in detail in \citet{Khadkaetal2023} and we list the main observables, i.e. the redshift $z$, the rest-frame \Mgii\ time delay $\tau$, 2 keV flux density per frequency $F_X$, 2500\,\AA\, flux density per frequency $F_{UV}$, and 3000\,\AA\, flux density $F_{3000}$, in Appendix~A of \citet{Khadkaetal2023}. The sample consists of 53 sources from the Sloan Digital Sky Survey RM programme \citep[SDSS-RM;][]{2016ApJ...818...30S, shen_2019, Homayouni2020}, NGC 4151 \citep{2006ApJ...647..901M}, and 4 sources from the OzDES RM programme \citep{Zhefu2021}. These sources are also detected in the X-ray domain and are listed in the XMM-Newton X-ray source catalog (4XMMDR11). The sample position on the sky to the Galactic plane and the Galactic centre \citep[Sgr~A*; ][]{2017FoPh...47..553E} is shown in Fig.~\ref{fig_quasars_dustmap}, which is colour-coded using the decadic logarithm of the $(B-V)$ colour index $\log{E_{B-V}}$ according to \citet{1998ApJ...500..525S}. 

\begin{table*}[tbh]
    \centering
    \caption{Selected properties of the main sample of X-ray detected reverberation-mapped QSOs (left column) and of the subsample of 21 sources (right column; see Sec.\ \ref{sec:data} for details). The luminosities are computed for the flat $\Lambda$CDM model with $\Omega_{\rm m0}=0.3$ and $H_0=70\,{\rm km\,s^{-1}\,Mpc^{-1}}$. \Mgii\, time delays are expressed in the rest frames of the sources.}
    \begin{tabular}{c|c|c}
    \hline
    \hline
    Property  & Main sample & Subsample  \\
    \hline
    Source number      &  58    &   21   \\
    redshift range &  (0.0041, 1.686)       &   (0.418, 1.587)   \\
    redshift (16, 50, 84) \% percentiles & (0.527, 0.990, 1.454)    &  (0.4810, 0.919, 1.3394) \\
    2 keV luminosity range [${\rm erg\,s^{-1}}$] & ($1.9 \times 10^{41}$, $6.1 \times 10^{44}$)  & ($1.6\times 10^{43}$, $1.5 \times 10^{44}$)\\
    2 keV luminosity median [${\rm erg\,s^{-1}}$] & $4.4 \times 10^{43}$ & $4.8 \times 10^{43}$ \\
    2500\,\AA\, luminosity range [${\rm erg\,s^{-1}}$] & ($1.2 \times 10^{43}$, $1.3 \times 10^{46}$)  & ($5.9 \times 10^{43}$, $4.8 \times 10^{45}$)\\
    2500\,\AA\, luminosity median [${\rm erg\,s^{-1}}$] & $9.3\times 10^{44}$  & $9.5\times 10^{44}$ \\
    \Mgii\, time delay range [days] & (5.3, 387.9)  & (17.2 , 387.9)  \\
    \Mgii\, time delay median [days] & 92.0  &  99.1 \\
    \hline 
    \end{tabular} 
    \label{tab_sample_description}
\end{table*}

We summarize the main properties of the main sample in Table~\ref{tab_sample_description}. 
In \citet{Khadkaetal2023} we also estimated the $\alpha_{OX}$ parameter, finding that our sample is not heavily obscured. An alternative verification that the \Mgii\ quasars are not intrinsically obscured sources (or red quasars) is presented in Appendix~\ref{appendix:reddening} using $ugriz$ magnitudes from the SDSS database.

In addition, we analyze a subsample of 21 sources that meet the hard X-ray index and far UV slope criteria of \citet{Lussoetal2020}.\footnote{The hard X-ray photon index should lie between 1.7 and 2.8 and the far-UV spectral slope should lie between $-0.7$ and 1.5 \citep{Lussoetal2020}. The extinction cuts are applied to the sources that are outside these limits.} The selection methodology and the subsample was described in detail in \citet{Khadkaetal2023}.\footnote{Using the source identification (ID) of \citet{Khadkaetal2023} (see their Appendix A), the IDs of the sources are the following: 18, 28, 44, 118, 159, 185, 260, 280, 301, 303, 338, 440, 449, 459, 522, 588, 675, J141214, J141018, J141650, and J141644.} The properties of the subsample are also described in Table~\ref{tab_sample_description}.

\section{Cosmological models and parameters}
\label{sec:models}

The application of these QSO observations in cosmology depends on the empirical measurement of the QSO luminosity distances. This requires the assumption of a cosmological model. However, to determine whether or not the QSOs are standardizable requires that we study them in several different cosmological models to see whether or not the empirical correlation relation used to determine their luminosities is independent of the assumed cosmological model \citep{KhadkaRatra2020c}. In this paper, we use three spatially-flat and three spatially non-flat\footnote{For discussions of constraints on spatial curvature see \citet{Ranaetal2017}, \citet{Oobaetal2018b}, \citet{ParkRatra2019b}, \citet{DESCollaboration2019}, \citet{EfstathiouGratton2020}, \citet{DiValentinoetal2021a}, \citet{Khadkaetal2021}, \citet{Dhawanetal2021}, \citet{2021PDU....3300851V,2021ApJ...908...84V}, \citet{Renzietal2021}, \citet{Gengetal2022}, \citet{CaoKhadkaRatra2022}, \citet{MukherjeeBanerjee2022}, \citet{Glanvilleetal2022}, \citet{Wuetal2022}, \citet{deCruzPerez2022}, \citet{DahiyaJain2022}, \citet{Stevensetal2022}, \cite{Favaleetal2023}, and references therein.} cosmological models to determine QSO luminosity distances. In any cosmological model, the luminosity distance can be computed as a function of redshift ($z$) and cosmological parameters ($\mathbf{p}$) in the following way,
\begin{equation}
  \label{eq:DL}
\resizebox{0.475\textwidth}{!}{%
    $D_L(z,\mathbf{p}) = 
    \begin{cases}
    \frac{c(1+z)}{H_0\sqrt{\Omega_{k0}}}\sinh\left[\frac{H_0\sqrt{\Omega_{k0}}}{c}
    D_C(z,\mathbf{p})\right] & \text{if}\ \Omega_{k0} > 0, \\
    \vspace{1mm}
    (1+z)D_C(z,\mathbf{p}) & \text{if}\ \Omega_{k0} = 0,\\
    \vspace{1mm}
    \frac{c(1+z)}{H_0\sqrt{|\Omega_{k0}|}}\sin\left[\frac{H_0\sqrt{|\Omega_{k0}|}}{c}
    D_C(z,\mathbf{p})\right] & \text{if}\ \Omega_{k0} < 0.
    \end{cases}$%
    }
\end{equation}
Here $c$ is the speed of light, $H_0$ is the Hubble constant, $\Omega_{k0}$ is the current value of the spatial curvature energy density parameter, and $D_C(z,\mathbf{p})$ is the comoving distance.  $D_C(z,\mathbf{p})$ is computed as a function of $z$ and $\mathbf{p}$ for a given cosmological model as follows,
\begin{equation}
\label{eq:gz}
   D_C(z,\mathbf{p}) = c\int^z_0 \frac{dz'}{H(z',\mathbf{p})},
\end{equation}
where $H(z,\mathbf{p})$ is the Hubble parameter that is given next for the six cosmological models we use in this paper.

In a compact form, the Hubble parameter for all the six models (spatially flat/non-flat $\Lambda$CDM model, XCDM parametrization, and $\phi$CDM model) can be written as
\begin{equation}
\label{eq:friedLCDM}
    H(z) = H_0\sqrt{\Omega_{m0}(1 + z)^3 + \Omega_{k0}(1 + z)^2 + \Omega_{DE}(z)},
\end{equation}
where $\Omega_{k0}$ vanishes in the spatially-flat models. For the $\Lambda$CDM models and XCDM parametrizations, $\Omega_{DE}(z)$ = $\Omega_{DE0}(1+z)^{1+\omega_X}$, where $\Omega_{DE0}$ is the current value of the dark energy density parameter and $\omega_X$ is the dark energy equation of state parameter. For the $\Lambda$CDM models $\omega_X = -1$ and the $X$CDM parametrizations $\omega_X$ is a free parameter to be determined from observational data. For the $\phi$CDM models \citep{PeeblesRatra1988, RatraPeebles1988, Pavlovetal2013}, $\Omega_{DE}(z)$ = $\Omega_{\phi}(z, \alpha)$ can be obtained by solving the equations of motion of a spatially homogeneous scalar field model numerically. Here $\alpha$ is a positive parameter characterizing the inverse-power-law potential energy density of the dynamical dark energy scalar field ($\phi$) and can be constrained by using observational data.\footnote{For discussions of constraints on the $\phi$CDM model see \citet{Zhaietal2017}, \citet{Oobaetal2018c, Oobaetal2019}, \citet{ParkRatra2018, ParkRatra2019c, ParkRatra2020}, \citet{SolaPercaulaetal2019}, \citet{Singhetal2019}, \citet{UrenaLopezRoy2020}, \citet{SinhaBanerjee2021}, \citet{Xuetal2021}, \citet{deCruzetal2021}, \citet{Jesusetal2021}, \citet{Adiletal2022}, \citet{Dongetal2023}, \citet{VanRaamsdonkWaddell2023}, \citet{CaoRatra2023b} references therein.}

To determine QSO luminosity distances using empirical correlation relations, in particular, those expressed by Eqs.\ (\ref{eq_R-L}) and (\ref{eq_UV-Xray}) below, we perform likelihood analysis of predicted luminosity distances in a given cosmological model using Eq.\ (\ref{eq:DL}) and the observational luminosity distances obtained from Eq.\  (\ref{eq_DL_R-L}) or Eq.\ (\ref{eq_DL_LX-LUV}) below. This allows us to measure the nuisance parameters involved in the correlation relation (and check whether they are independent of cosmological model parameters) and that ultimately leads us to the determination of the QSO luminosity distances. For a detailed description of the computational method and the determination of QSO luminosity distances used in this paper, see \cite{Khadkaetal2023}.

\section{UV and X-ray Extinction and luminosity distances}
\label{sec:methods}

The source intrinsic UV and X-ray flux densities (per frequency) at 2500\,\AA\, and 2 keV are $F_{\rm UV, int}$ and $F_{\rm X,int}$, respectively, originate in the very central parts of galaxies during the accretion process and as such are assumed not to be affected by dust at the place of their origin. The corresponding luminosities are calculated as $L_{\rm UV,int}=4\pi D_{\rm L}^2 F_{\rm UV, int}$ and $L_{\rm X,int}=4\pi D_{\rm L}^2 F_{\rm X,int}$, where $D_{\rm L}$ is the luminosity distance of a given QSO. 

For the $R-L$ relation, the corresponding UV luminosity is expressed at $3000\,$\AA, $L_{\rm 3000,int}=4\pi D_{\rm L}^2 F_{3000,\nu}\nu_{3000}=4\pi D_{\rm L}^2 F_{\rm UV, int}(2500/3000)^{\alpha_{\nu}}\nu_{3000}$, where $\alpha_{\nu}\simeq -0.45$ is the mean QSO continuum slope in the frequency domain \citep[$F_{\nu}\propto \nu^{\alpha_{\nu}}$;][]{2001AJ....122..549V} and $\nu_{3000}$ is the frequency corresponding to 3000\,\AA. Since the relation between the mean radius of the BLR region $R$ and the corresponding time delay $\tau$ in the rest frame of the source is given by the light-travel relation, $R=c\tau$, the $R-L$ relation can be expressed in the form using $\tau$ instead of $R$,
\begin{equation}
    \log{\left(\frac{\tau}{{\rm days}}\right)}=\beta+\gamma \log{\left(\frac{L_{\rm 3000,int}}{10^{\eta}\,{\rm erg\,s^{-1}}} \right)}\,
    \label{eq_R-L}
\end{equation}
where $\gamma$, $\beta$, and $\eta$ represent the slope, intercept, and normalization coefficients, respectively. From Eq.~\eqref{eq_R-L}, we can derive an expression for the luminosity distance $D_{L,R-L}$ as a function of $\tau$, $F_{\rm UV, int}$, and the coefficients of the $R-L$ relation,
 \begin{align}
 \label{eq_DL_R-L}
    &\log{D_{L,R-L}}=\, \\
    &\frac{1}{2\gamma}\{\log{\tau}-\beta -\gamma[\log{(4\pi)}+\log{F_{\rm UV, int}}+15.04-\eta]\}\,, \notag
\end{align}
where the term $15.04$ results from the evaluation of $\log{[(2500/3000)^{\alpha_{\nu}}\nu_{3000}]}$.
The $L_{X}-L_{UV}$ relation considering the intrinsic X-ray and UV luminosities of the source $L_{\rm X,int}$ and $L_{\rm UV,int}$ located at the luminosity distance $D_{L,L_{X}-L_{UV}}$ can be expressed as
\begin{equation}
    \log{\left(\frac{L_{\rm X,int}}{{\rm erg\,s^{-1}\,Hz^{-1}}}\right)}=\beta'+\gamma'\log{\left(\frac{L_{\rm UV,int}}{10^{\eta'}\,{\rm erg\,s^{-1}Hz^{-1}}}\right)}\,,
    \label{eq_UV-Xray}
\end{equation}
where $\beta'$, $\gamma'$, and $\eta'$ denote quantities analogous to those in Eq.~\eqref{eq_R-L}. The luminosity distance $D_{L,L_{X}-L_{UV}}$ inferred from the $L_{X}-L_{UV}$ relation given in Eq.~\eqref{eq_UV-Xray} is,
\begin{align}
\label{eq_DL_LX-LUV}
    &\log{D_{L,L_{X}-L_{UV}}}=\,\\
    &\frac{1}{2(1-\gamma')}[\beta'+(\gamma'-1)\log{(4\pi)}+\notag\\
    &\gamma'(\log{F_{\rm UV, int}}-\eta')-\log{F_{\rm X,int}}]\,. \notag 
\end{align}

Taking into account just the intrinsic UV and X-ray flux densities, we can evaluate the luminosity-distance difference, $\Delta \log{D_{\rm L}}$, as
\begin{align}
   &(\log{D_{L,L_{X}-L_{UV}}}-\log{D_{L,R-L}})_{\rm int}=\\
   &\delta+\frac{\log{F_{\rm UV, int}}-\log{F_{\rm X,int}}}{2(1-\gamma')}\,,\notag
   \label{eq_DL_diff}
\end{align}
where the factor $\delta$ is a function of $\gamma$, $\beta$, $\gamma'$, $\beta'$, and $\tau$,
\begin{equation}
   \delta=\frac{\beta-\log{\tau}}{2\gamma}+\frac{\beta'-\gamma'\eta'}{2(1-\gamma')}+7.52-\frac{\eta}{2}\,,  
\end{equation}
where the term 7.52 results from the evaluation of $0.5\log{[(2500/3000)^{\alpha_{\nu}}\nu_{3000}]}$.

In the following, we adopt the assumption that, in the absence of a systematic effect, the luminosity distance difference for a given source, or the statistical mean (median) of the luminosity-distance differences for a source sample, vanish, i.e. $(\Delta \log{D_{\rm L}})_{\rm int}\equiv 0$, hence $\delta+(\log{F_{\rm UV, int}}-\log{F_{\rm X,int}})/[2(1-\gamma')]=0$. However, during the propagation of light through the circumnuclear and interstellar medium of the host galaxy, both UV and X-ray photons are absorbed and scattered with wavelength-dependent optical depths $\tau_{UV}$ and $\tau_{X}$, respectively. The observed flux densities are therefore attenuated following the exponential law, $F_{UV}=F_{\rm UV, int}\exp{(-\tau_{UV})}$ and $F_{X}=F_{\rm X,int}\exp{(-\tau_{X})}$. Hence, the resulting luminosity-distance difference under the influence of extinction in both the UV and X-ray domains is,
\begin{align}
\label{eq_DL_diff_ext}
    &(\Delta \log{D_{\rm L}})_{\rm ext}=\,\\
    &\delta+\frac{\log{F_{\rm UV, int}}-\log{F_{\rm X,int}}}{2(1-\gamma')}+\frac{(\tau_{X}-\tau_{UV})\log{e}}{2(1-\gamma')}=\notag\\
    &\frac{(\tau_{X}-\tau_{UV})\log{e}}{2(1-\gamma')} \,, \notag  
\end{align}
where the second equation follows because the sum of the first two terms on the right-hand side of the first of Eqs.~\eqref{eq_DL_diff_ext} is intrinsically zero. In Eq.~\eqref{eq_DL_diff_ext}, $\log{e}$ denotes the decadic logarithm of Euler's number, i.e. $\log{e}\simeq 0.434$. We outline a more detailed, step-by-step derivation of Eq.~\eqref{eq_DL_diff_ext} in Appendix~\ref{appendix: formulae}. The relation among the distributions of $\delta$, $(\log{F_{UV}}-\log{F_{X}})/[2(1-\gamma')]$, and $\Delta \log{D_{\rm L}}$ for our sample is discussed in Appendix~\ref{appendix: delta-flux}. 

Consequently, the optical depth difference $\tau_{X}-\tau_{UV}$ can be expressed as a function of the luminosity-distance difference and the slope $\gamma'$ of the $L_{X}-L_{UV}$ relation,
\begin{equation}
    \tau_{X}-\tau_{UV}=\frac{2(1-\gamma')}{\log{e}}(\Delta \log{D_{\rm L}})_{\rm ext}\,.    
\end{equation}
The extinction in magnitudes at a given wavelength is directly proportional to the optical depth, $A_{\lambda}=1.086\tau_{\lambda}$. Hence we can express the X-ray/UV colour excess as
\begin{align}
    &E_{X-UV}\equiv A_{X}-A_{UV}\simeq \,  \label{eq_Exuv_DeltaDL}\\
    &5.001(1-\gamma')\left\langle(\Delta \log{D_{\rm L}})_{\rm ext}\right\rangle\,,\notag
\end{align}
where the angular brackets represent the median/mean/peak value of the luminosity-distance difference for a given population of QSOs.

\begin{figure*}
    \centering
    \includegraphics[width=0.9\columnwidth]{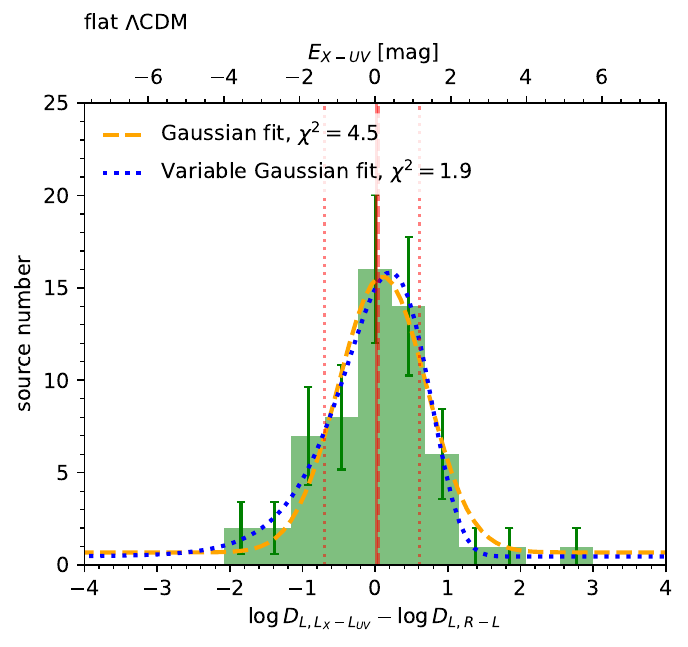}
    \includegraphics[width=0.9\columnwidth]{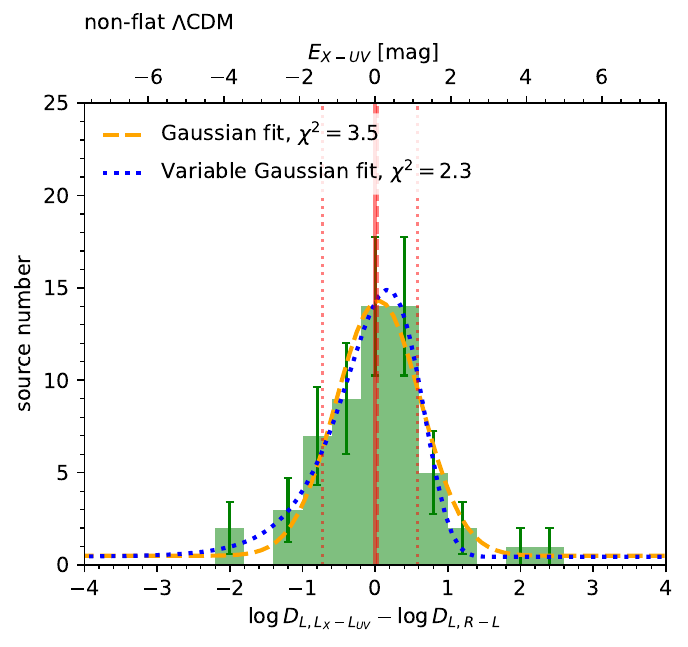}
     \includegraphics[width=0.9\columnwidth]{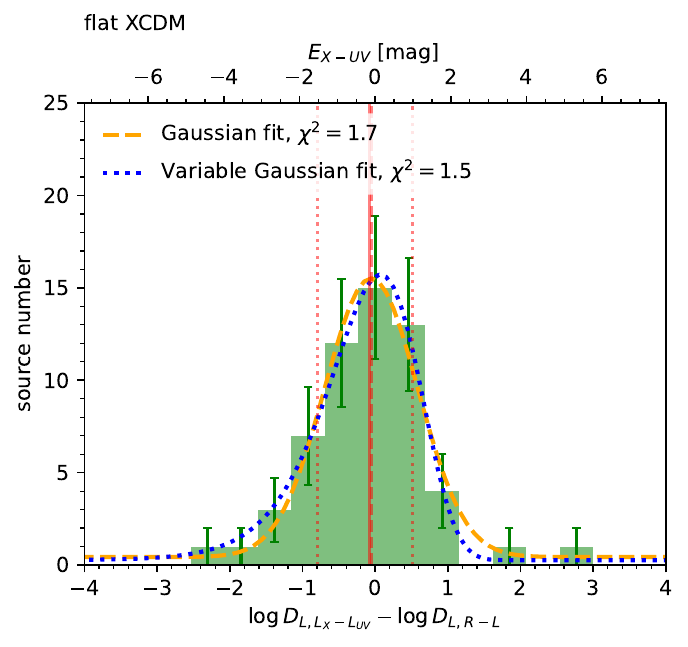}
    \includegraphics[width=0.9\columnwidth]{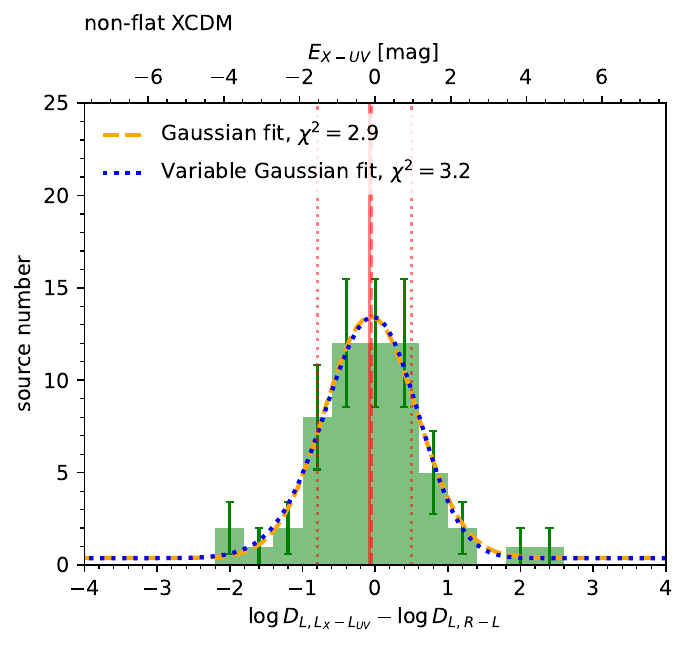}
    \includegraphics[width=0.9\columnwidth]{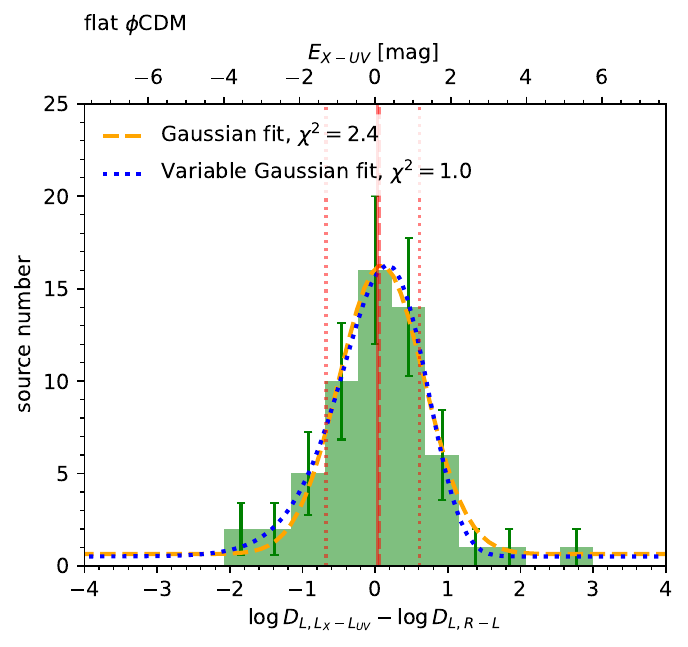}
    \includegraphics[width=0.9\columnwidth]{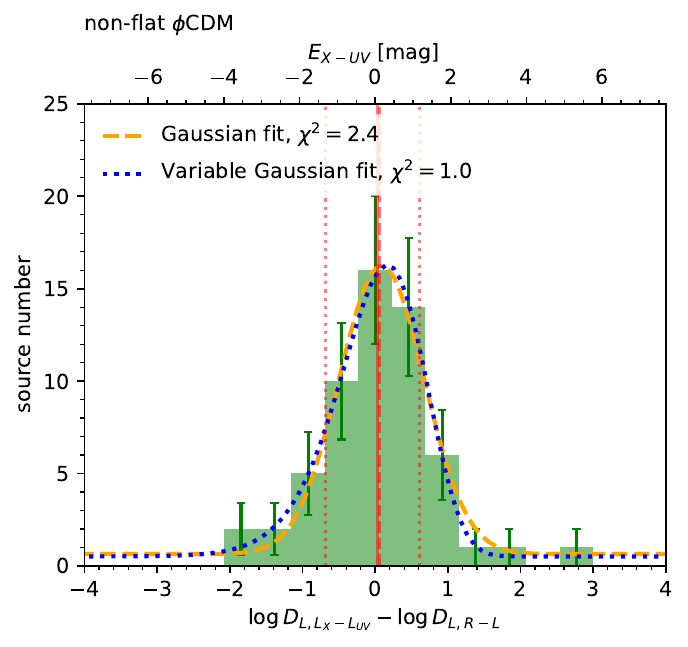} 
    \caption{Distributions of the quasar luminosity-distance differences $\Delta \log{D_{\rm L}}=\log{D_{L,L_X-L_{UV}}}-\log{D_{L,R-L}}$ for 58 sources for flat and non-flat $\Lambda$CDM, XCDM, and $\phi$CDM cosmological models (from top to bottom row). The X-ray/UV colour index $E_{X-UV}=5.001(1-\gamma')\Delta \log{D_{\rm L}}$ is enumerated along the top $x$ axis in each panel. Solid red vertical lines are the difference means, dashed red vertical lines are the difference medians, and dotted red vertical lines are $16\%$ and $84\%$ percentiles. The bin width is determined based on the Knuth binning algorithm and the source-number uncertainties for each bin are $\sigma_{y,i}=\sqrt{N_i}$ where $N_i$ is the number of points in each bin. The orange dashed line depicts the best-fit Gaussian function and the blue dotted line shows the best-fit variable Gaussian function.}
    \label{fig_DL_diff}
\end{figure*}

\section{Results}
\label{sec:results}

 In \citet{Khadkaetal2023} we simultaneously measured the $R-L$ or $L_{X}-L_{UV}$ relation parameters and the cosmological-model parameters for six different cosmological models -- flat and non-flat $\Lambda$CDM, XCDM, and $\phi$CDM. Based on the cosmological parameter values, we computed the $R-L$-based and $L_{X}-L_{UV}$-based luminosity distances, $D_{L,R-L}$ and $D_{L,L_{X}-L_{UV}}$, respectively. In Fig.~\ref{fig_DL_diff} we show the distributions of $\Delta \log{D_{\rm L}}$. The plot shows that the difference between the luminosity distances can be positive or negative. If extinction plays the dominant role in measurement accuracy, the difference should always be positive. The presence of negative values implies that other measurement errors dominate. On the other hand, if no extinction effect is present, the mean/median values should be consistent with zero. Establishing the systematic shift is important, particularly from the point of view of future measurements coming from very large samples, when the stochastic net error for the entire sample will become small but the systematic shift will persist. With this aim, we concentrate on tracing this systematic shift in the present sample.
 
 In comparison with \citet{Khadkaetal2023}, here we construct distributions of non-normalized luminosity-distance differences, from which $E_{X-UV}$ can be inferred using Eq.~\eqref{eq_Exuv_DeltaDL}. The normalized luminosity-distance distributions computed in \citet{Khadkaetal2023} are appropriate for comparing the $R-L$-based and $L_{X}-L_{UV}$-based luminosity distances for each source. For the graphical representation in Fig.~\ref{fig_DL_diff}, we binned $\Delta \log{D_{\rm L}}$ using Knuth's rule \citep{2006physics...5197K}. The uncertainty in the number of sources falling into the bin is estimated as $\sigma_{y,i}=\sqrt{N_i}$.

\begin{table*}
    \centering
     \caption{Characteristics of the distributions $\Delta \log{D_{\rm L}}=\log{D_{L,L_{X}-L_{UV}}}-\log{D_{L,R-L}}$ for six cosmological models listed in the first column. The distributions are graphically depicted in Fig.~\ref{fig_DL_diff}. From the left to the right columns, we list the distribution median, 16\%- and 84\%-percentiles, mean, skewness and Fisher's kurtosis parameters (both corrected for statistical bias), the kurtosis test, and the Kolmogorov-Smirnov (KS) test statistic including the corresponding $p$-values.}
    \begin{tabular}{c|c|c|c|c|c|c|c|c}
    \hline
    \hline
    Model & Median & 16\% & 84\% & Mean & Skewness & Fisher's kurtosis & Kurtosis test & KS test\\
    \hline
    flat $\Lambda$CDM & 0.0423 & $-0.7036$ & 0.6080  & 0.0167  & 0.2369 & 1.7366 & 2.1136, $p=0.035$ & 0.0000, $p=1.0000$ \\
    non-flat $\Lambda$CDM & 0.0310 & $-0.7229$  & 0.5806 & $-0.0030$  & 0.2194 &  1.7271 & 2.1068, $p=0.035$  &0.0690, $p=0.9993$  \\
    flat XCDM  &  $-0.0539$  & $-0.7911$ & 0.5106 & $-0.0768$  & 0.2418 & 1.7391 & 2.1154, $p=0.034$ & 0.1034, $p=0.9192$ \\
    non-flat XCDM & $-0.0490$  & $-0.7969$ & 0.5003 & $-0.0812$  & 0.2228 & 1.7290 & 2.1082, $p=0.035$ & 0.1034, $p=0.9192$ \\ 
    flat $\phi$CDM & 0.0534 & $-0.6766$ & 0.6097 & 0.0293  & 0.2390 & 1.7376 & 2.1144, $p=0.034$ & 0.0345, $p=1.0000$ \\
    non-flat $\phi$CDM & 0.0556  & $-0.6780$ & 0.6128 &  0.0307  & 0.2375 &  1.7369 & 2.1139, $p=0.035$ &0.0517, $p=1.0000$  \\
    \hline
    flat $\Lambda$CDM - 21 & 0.1869 & $-0.4825$ & 0.5173 & 0.0114  & $-0.8175$ & $0.2430$& 0.4899, $p=0.624$ & $0.1552$, $p=0.7855$  \\ 
    \hline 
    \end{tabular}   
    \label{tab_diff_dist}
\end{table*}

The distributions of the luminosity distance differences have several common characteristics. At first glance, the peak of each distribution is shifted to the positive side, which hints at positive values of $\tau_{X}-\tau_{UV}$ and hence also $E_{X-UV}$. We summarize the main characteristics, specifically the distribution median, 16\%, and 84\% percentiles, mean, skewness, Fisher's kurtosis, kurtosis test, and the Kolmogorov-Smirnov test statistic, in Table~\ref{tab_diff_dist}. More specifically, for calculating the distribution skewness, we correct for the statistical bias and use the Fisher-Pearson standardized moment $\sqrt{N(N-1)}/(N-2) m_3/m_2^{3/2}$, where $N$ is the sample size and $m_2$ and $m_3$ denote the second and the third central moments, respectively. All of the distributions are positively skewed, which is caused by the presence of tails on the positive side. For four out of six models, the median is positive; it is negative for the flat and non-flat XCDM models. The mean value is positive for flat $\Lambda$CDM, flat and non-flat $\phi$CDM models, while for the other cosmological models, it is negative. Since the mean is more sensitive to the outliers in the distribution tails, we consider the median values to be more representative of the quasar sample. Fisher's kurtosis, which is also corrected for the statistical bias, is greater than zero, which implies a heavier tail than for the normal distribution. We verify the deviation from the normal distribution by performing the kurtosis test, whose $z$-scores and the corresponding $p$-values confirm the deviation. The two-sample Kolmogorov-Smirnov (KS) test is applied between each model and the flat $\Lambda$CDM model. All of the KS $p$-values are close to one, hence the null hypothesis that the $\Delta \log{D_{\rm L}}$ distributions are drawn from the same underlying distribution as the one for the flat $\Lambda$CDM model is valid.

\begin{table*}
    \centering
    \caption{Parameters of the X-ray/UV extinction $E_{X-UV}$  in magnitudes for six cosmological models inferred using the variable Gaussian function that fits the $E_{X-UV}$ distributions better than a normal Gaussian function. We transform the luminosity-distance difference $\Delta \log{D_{\rm L}}$ to the colour excess $E_{X-UV}$ using Eq.~\eqref{eq_Exuv_DeltaDL} with the best-fit slope of the $L_{X}-L_{UV}$ relation (second column) adopted from \citet{Khadkaetal2023}. The third, fourth, fifth, and sixth columns (from the left to the right) list $E_{X-UV}$ median and peak values, the standard deviations to the positive and the negative sides of the distributions, respectively. The quoted errors are the errors of the fit of the variable Gaussian function to the $E_{X-UV}$ histograms.}
    \begin{tabular}{c|c|c|c|c|c}
    \hline
    \hline
    Model  & $\gamma'$ &  $E_{X-UV}$ [mag] (median) & $E_{X-UV}$ [mag] (peak) & $\sigma_{+}$ [mag] & $\sigma_{-}$ [mag]\\
    \hline
    flat $\Lambda$CDM    & $0.616 \pm 0.074$ & 0.081 & $0.29 \pm 0.13$ & $1.06 \pm 0.11$   & $1.29 \pm 0.12$ \\
    non-flat $\Lambda$CDM     &$0.609 \pm 0.073$ & 0.061 & $0.31 \pm 0.10$ & $0.95 \pm 0.09$ & $1.25 \pm 0.10$ \\    
    flat XCDM                 &$0.614 \pm 0.075$    & -0.104 & $0.17 \pm 0.13$ & $1.02 \pm 0.12$  & $1.29 \pm 0.13$  \\ 
    non-flat XCDM            &$0.608 \pm 0.075$    & -0.096 & $0.20 \pm 0.11$& $0.89 \pm 0.10$   & $1.20 \pm 0.11$ \\
    flat $\phi$CDM             &$0.609 \pm 0.073$   & 0.104 & $0.34 \pm 0.11$ & $1.03 \pm 0.10$   & $1.28 \pm 0.11$ \\    
    non-flat $\phi$CDM        &$0.610 \pm 0.073$    & 0.108 &  $0.34 \pm 0.11$ & $1.03 \pm 0.10$   &  $1.28 \pm 0.11$ \\ 
    \hline
    flat $\Lambda$CDM - 21  & $0.610 \pm 0.100$   & 0.364 &  $0.50 \pm 0.12$& $0.82 \pm 0.09$  &  $1.20 \pm 0.13$ \\
    \hline       
    \end{tabular}  
    \label{tab_E_X_UV}
\end{table*}

Furthermore, we analyze the $\Delta \log{D_{\rm L}}$ distributions in Fig.~\ref{fig_DL_diff} by fitting a Gaussian function and a variable Gaussian function to them.\footnote{For the variable Gaussian function, we use the form according to \citet{2004physics...6120B}. The variable Gaussian function is introduced as $G_{\rm var}=a\exp{(-(x-x_0)^2/[2\sigma(x)])}+b$, where $\sigma(x)=\sigma_1+\sigma_2(x-x_0)$, $\sigma_1=(2\sigma_{+}\sigma_{-})/(\sigma_{+}-\sigma_{-})$, and $\sigma_2=(\sigma_{+}-\sigma_{-})/(\sigma_{+}+\sigma_{-})$.} The comparison of the fits based on the $\chi^2$ value shows that the variable Gaussian function fits the distributions better, with the peak value shifted to the positive side and with $\sigma_{-}>\sigma_{+}$, i.e. the distributions are asymmetric to the peak. Qualitatively the same behaviour is traced for the distributions of the X-ray/UV colour index $E_{X-UV}$ (in magnitudes), which is calculated using Eq.~\eqref{eq_Exuv_DeltaDL} and its value is greater than the luminosity-distance offset by about a factor of two. We show the corresponding histograms binned according to Knuth's rule, including the Gaussian and the variable Gaussian fits, for all the flat and non-flat $\Lambda$CDM, XCDM, and $\phi$CDM cosmological models in Fig.~\ref{fig_E_xuv} in Appendix~\ref{appendix: E_X_UV}. In Fig.~\ref{fig_E_xuv}, the distribution means are depicted with red solid vertical lines, medians are represented by red vertical dashed lines, and the 16\%- and 84\%-percentiles are shown with red vertical dotted lines. The basic statistical properties of these distributions, specifically the $E_{X-UV}$ median and peak values, the right and the left standard deviations, $\sigma_{+}$ and $\sigma_{-}$, respectively, are summarized in Table~\ref{tab_E_X_UV}. For all the cosmological models, we obtain $\sigma_{+}<\sigma_{-}$, which implies a significant asymmetry. The median values are predominantly positive (except for flat and non-flat XCDM models), with an average value of $\overline{E}_{X-UV}=0.03$ mag, which is comparable in magnitude to the value inferred from the mean for the flat $\Lambda$CDM model. The inferred peak values of $E_{X-UV}$ are positive for all the cosmological models. The average value is $\overline{E}_{X-UV}=0.28 \pm 0.07$ mag, hence larger than the value inferred from the medians. Overall, based on the average median and the peak values, we estimate the X-ray/UV colour index of $\overline{E}_{X-UV}\sim 0.03-0.28$ mag for our sample.

The positive peak value of the $E_{X-UV}$ distributions for all the cosmological models implies that the extinction is present in the sample, and the effect is stronger in the X-ray band than in the UV band. As we show by the comparison of distributions of $\delta$ and $(\log{F_{UV}}-\log{F_{X}})/[2(1-\gamma')]$ in Appendix~\ref{appendix: delta-flux}, the positive difference of $\tau_{X}-\tau_{UV}$ results in the shift of their sum to positive values due to the extinction term. In other words, extinction causes the drop of the X-ray flux to the UV flux density, i.e. $\log{F_{UV}}-\log{F_{X}}$ is positive, which is correlated with the positive difference in optical depths, $\tau_{X}-\tau_{UV}$. If we assume zero extinction in the UV domain, we can convert $\overline{E}_{X-UV}=0.03-0.28$ mag to the hydrogen column density $N_H$ which is customarily used in X-ray studies. Assuming just electron scattering, we derive the mean intrinsic column density of $N_H = 4.2 \times 10^{22} - 3.9 \times 10^{23}$ cm$^{-2}$, which is moderate and in the Compton-thin regime. It can even be lower if the effect is partially due to X-ray absorption. If UV extinction is also present, then the corresponding $N_H$ would be higher since we determine only the difference between the two effects. However, it is not likely that the two effects just compensate, so both X-ray and UV extinction effects are noticeable but not dramatically strong in our sample.

Assuming that in our sample UV extinction is negligible, we can use the mean X-ray extinction determined from our sample to correct the measured values of the $\log{D_{L,L_{X}-L_{UV}}}$. All actual values should be larger by 0.032 if corrected for X-ray extinction. When used for the flat $\Lambda$CDM cosmology, it systematically pushes the best solution towards significantly higher values of $\Omega_{m0}$, roughly by 0.14, as estimated from the standard cosmology calculator for a median redshift of 0.99, as in our sample. As we show in the derivations in Appendix~\ref{appendix: formulae}, see Eqs.~\eqref{eq_DL_R-L_app2} and \eqref{eq_DL_UV-X-ray_app3}, the extinction affects $\log{D_{L,R-L}}$ by a factor of $+0.2\tau_{UV}$, while the effect is more pronounced for $\log{D_{L,L_{X}-L_{UV}}}$ which is modified by $+0.54\tau_{X}-0.33\tau_{UV}$. Hence for a negligible UV extinction, only the $L_{X}-L_{UV}$ luminosity distance is affected.

Our analysis result is just the difference in the extinction effect in X-ray and UV bands and we cannot correct the sample without additional spectral studies.

\section{Discussion}
\label{sec:discussion}

We have shown that the $\Delta \log{D_{\rm L}}$ and hence also $E_{X-UV}$ distributions are significantly asymmetric and their peaks are shifted to the positive side for all the cosmological models considered. We attribute this to the extinction in the X-ray and UV domains with the average value in the range $\overline{E}_{X-UV}=0.03-0.28$ mag based on the average median and peak values of the $E_{X-UV}$ distributions for all six cosmological models.

 It implies that the effect is not very strong but it is present, and it can bias the cosmological results. Our sample of 58 sources contains all sources for which we have luminosity distance measurements with both methods, and we did not apply any pre-selection aimed at removing extinction based on spectral studies of individual sources. By calculating $\alpha_{OX}$ index \citep{Khadkaetal2023} and considering the $ugriz$ magnitudes from the SDSS catalogue (see Appendix~\ref{appendix:reddening}), we can conclude that our sample is mostly not heavily obscured and shares properties with the normal (blue) quasar population.

Nevertheless, even though our sample is not heavily obscured, we test the effect of extinction for our sample. By applying X-ray and UV selection cuts, we arrive at a subsample of 21 sources, for which the heavily absorbed sources were removed. Furthermore, we connect $E_{X-UV}$ colour index to the more standard $E_{B-V}$ in the optical domain. 

\subsection{Subsample without absorbed sources}
\label{subsection_subsample}

To test the extinction effect further, we attempt to remove sources with larger extinction. In our original sample of 58 quasars \citep{Khadkaetal2023} no pre-selection was made since the sample of X-ray-detected RM quasars is already limited in size. In the new test here we apply the criteria of \citet{Lussoetal2020}, as described in \citet{Khadkaetal2023}, to remove absorbed sources based on available hard X-ray photon indices and far-UV slopes. According to \citet{Lussoetal2020}, the hard X-ray photon index should lie between 1.7 and 2.8, which is satisfied for 41 sources, and the far-UV slope should lie between $-0.7$ and 1.5, which is satisfied for 27 out of 31 sources for which GALEX EUV magnitudes are available. Combining both criteria yields a subsample of 21 sources, which supposedly does not contain heavily extincted quasars; see Section~\ref{sec:data} for the subsample description.

\begin{figure*}
    \includegraphics[width=0.47\textwidth]{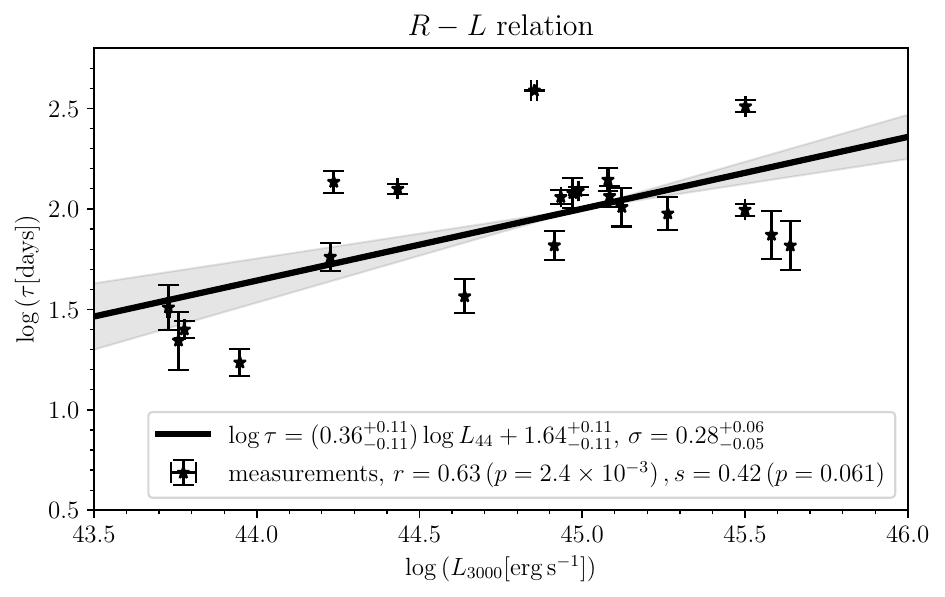}
    \includegraphics[width=0.47\textwidth]{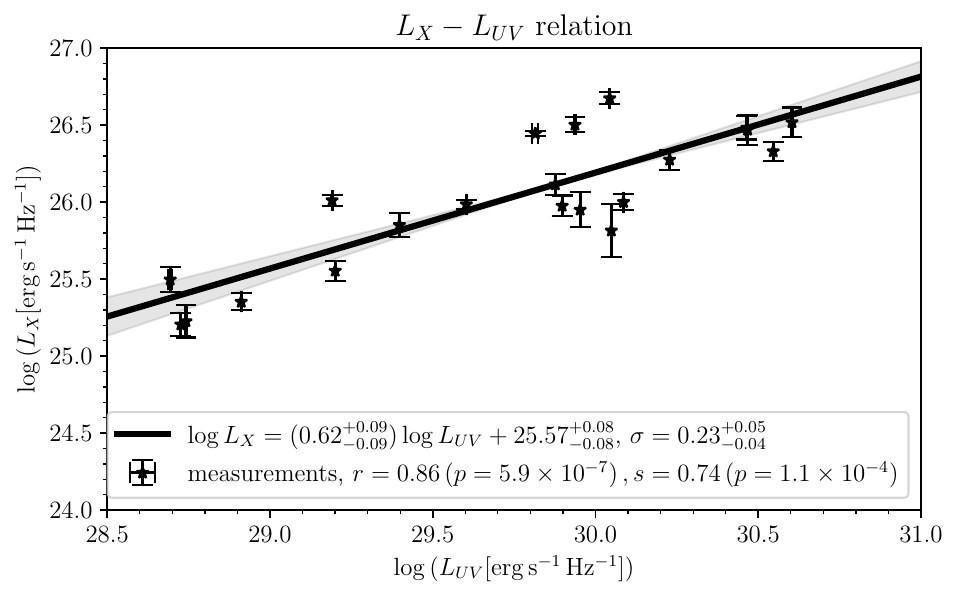}
    \includegraphics[width=0.47\textwidth]{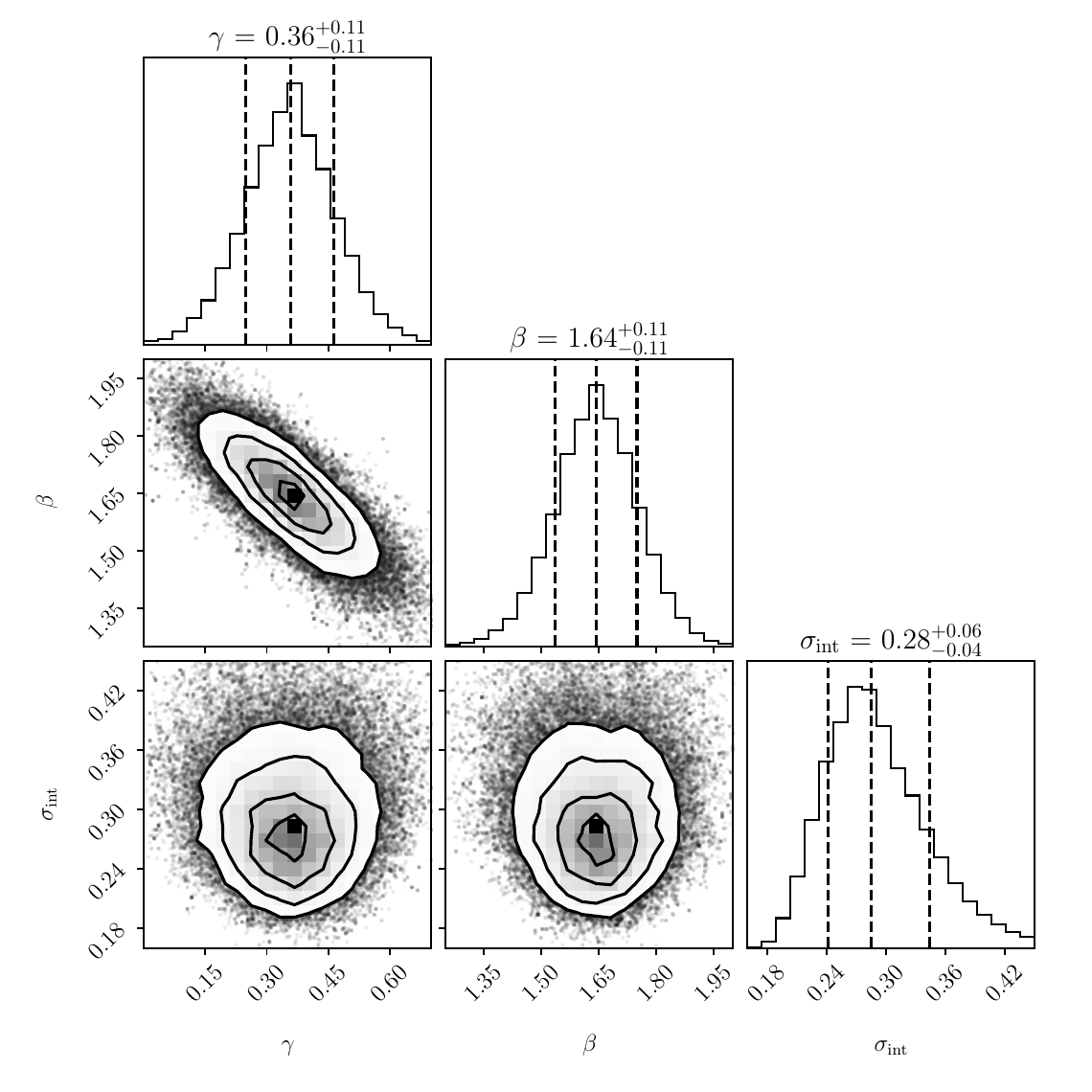}
    \hspace{0.5cm}
    \includegraphics[width=0.47\textwidth]{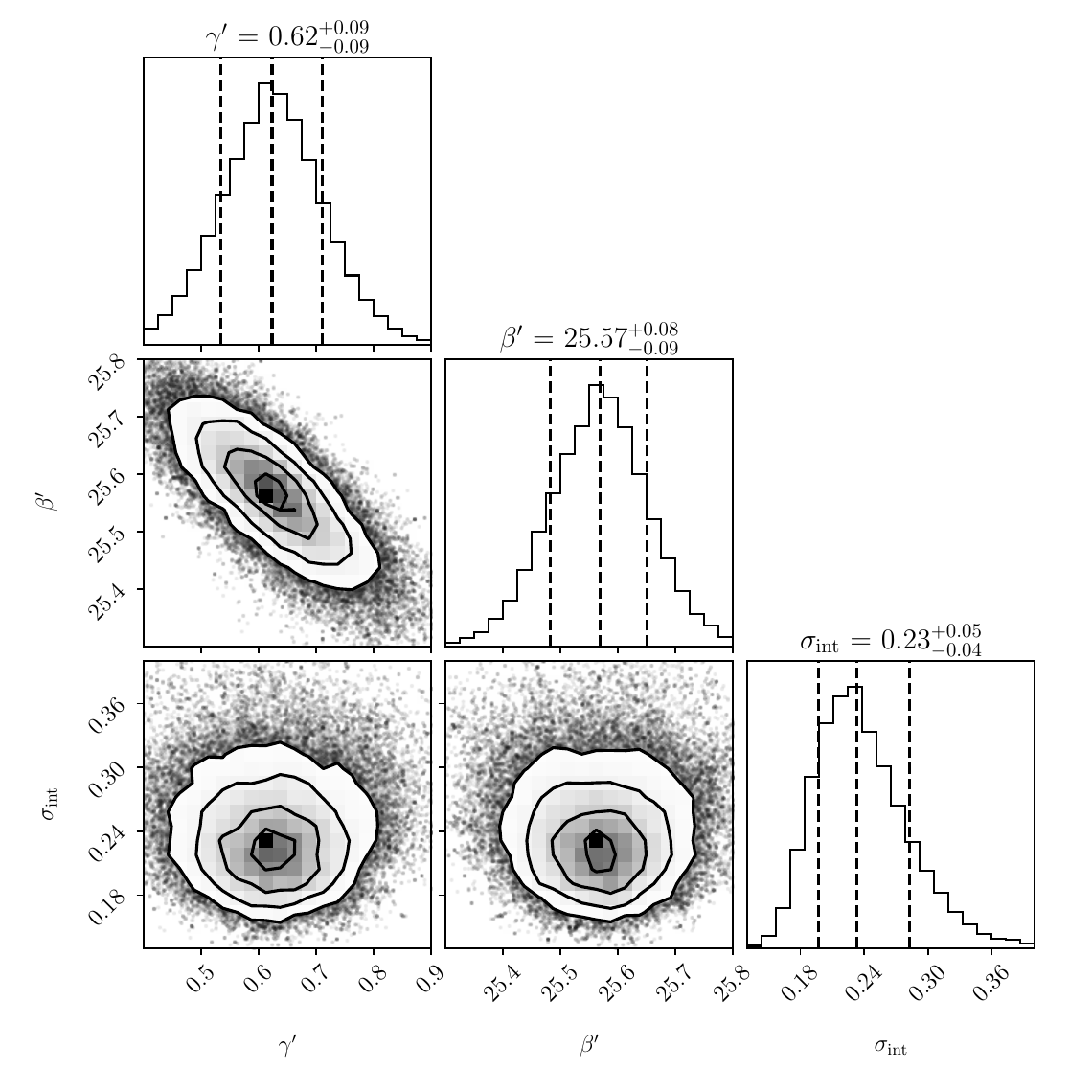} 
    \caption{$R-L$ and $L_{X}-L_{UV}$ relations in the flat $\Lambda$CDM model ($\Omega_{m0}=0.3$) in the left and the right panels, respectively, for the subsample of 21 sources. The top row depicts the measurements alongside the best-fit relations, including the intrinsic scatter $\sigma$, while the bottom row shows the likelihood distributions for the slopes $\gamma$ and $\gamma'$, the intercepts $\beta$ and $\beta'$, as well as the intrinsic scatter $\sigma_{\rm int}$ for the corresponding relations.}
    \label{fig_RL_LxLuv_21}
\end{figure*}

For these 21 sources, we construct $R-L$ and $L_{X}-L_{UV}$ relations in the flat $\Lambda$CDM model ($\om = 0.3$), see Fig.~\ref{fig_RL_LxLuv_21}, whose parameters are consistent with those for the corresponding relations in the main sample of 58 sources, with a slightly decreased intrinsic scatter of 0.28 and 0.23 dex for $R-L$ and $L_{X}-L_{UV}$ relations, respectively. The removal of absorbed sources is also beneficial for increasing the correlation in both relations. For the $R-L$ relation, we obtain a Pearson correlation coefficient of $r=0.63$ ($p=2.38 \times 10^{-3}$), while for the whole sample of 58 sources, it is $r = 0.56$ ($p = 4.42 \times 10^{-6}$). For the $L_{X}-L_{UV}$ relation of the subsample of 21 sources, we get $r=0.86$ ($p=5.90\times 10^{-7}$), while for the whole sample we have $r = 0.78$ ($p = 6.54 \times  10^{-13}$).

\begin{table*}
    \centering
    \caption{Marginalized one-dimensional best-fit parameters with 1$\sigma$ confidence intervals for the 21 $L_X-L_{UV}$ and $R-L$ QSOs in the flat $\Lambda$CDM model}
    \begin{tabular}{c|c|c|c|c|c}
    \hline
    \hline
    Model & Data & $\Omega_{m0}$ & $\sigma_{\rm int}$ & $\beta$, $\beta'$ & $\gamma$, $\gamma'$\\
    \hline
    Flat $\Lambda$CDM & $R-L$ QSOs & --   & $0.301^{+0.041}_{-0.069}$    & $1.660\pm0.120$ & $0.360\pm0.120$ \\
    & $L_X-L_{UV}$ QSOs & --   & $0.244^{+0.031}_{-0.054}$ & $25.541\pm0.090$ & $0.610\pm0.100$ \\
    \hline       
    \end{tabular}  
    \label{tab_relations_21}
\end{table*}

We also performed the simultaneous fitting of the $L_{X}-L_{UV}$ or $R-L$ relation parameters as well as cosmological parameters for the flat $\Lambda$CDM model. For both relations, we list the parameters in Table~\ref{tab_relations_21} including their 1$\sigma$ confidence intervals. $R-L$ and $L_{\rm X}-L_{\rm UV}$ relation parameters for 21 and 58 sources are consistent within the uncertainties. The differences in $\gamma$ and $\beta$ for the $R-L$ relation are $0.62\sigma$ and $0.15\sigma$, respectively. The differences in $\gamma'$ and $\beta'$ for the $L_{X}-L_{UV}$ relation are $0.04\sigma$ and $0.88\sigma$, respectively. 

\begin{figure}
    \centering
    \includegraphics[width=0.5\textwidth]{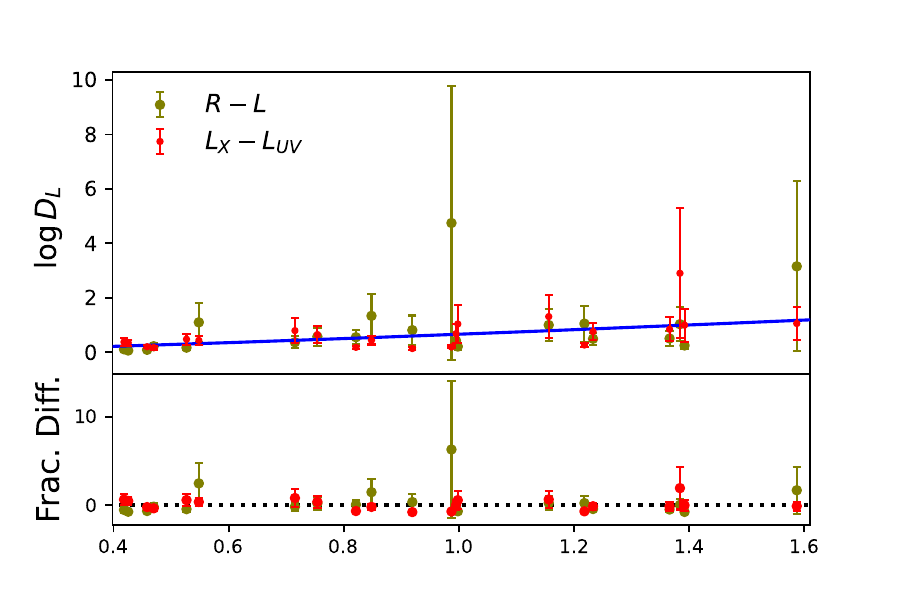}
    \caption{Upper panel: Luminosity distances for 21 sources (expressed in $10^4$ Mpc) inferred using $R-L$ and $L_{X}-L_{UV}$ relations in the flat $\Lambda$CDM model. The blue solid line is the prediction for the flat \lcdm\ model with $\om = 0.3$. Lower panel: Fractional difference between observed and model predicted luminosity distances for 21 sources. Colour follows the upper panel convention. The black dotted line indicates zero fractional difference.}
    \label{fig_lum_distance}
\end{figure}

\begin{figure*}
    \centering
    \includegraphics[width=0.43\textwidth]{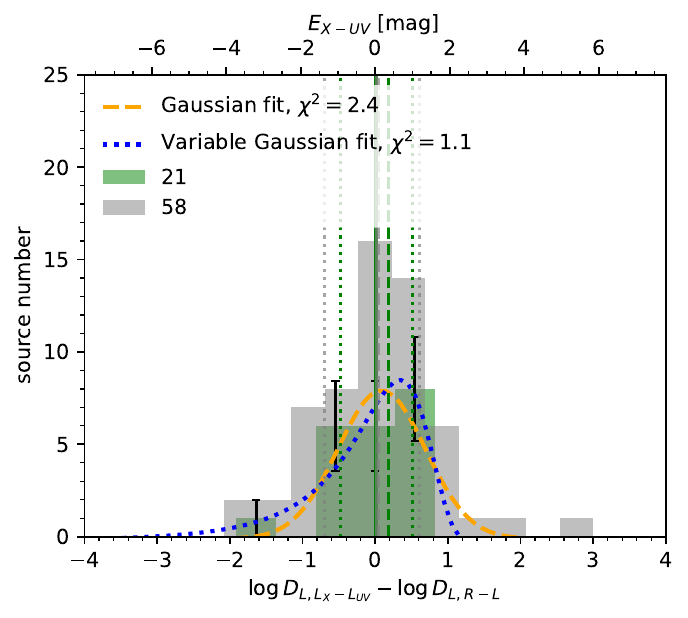}
     \includegraphics[width=0.44\textwidth]{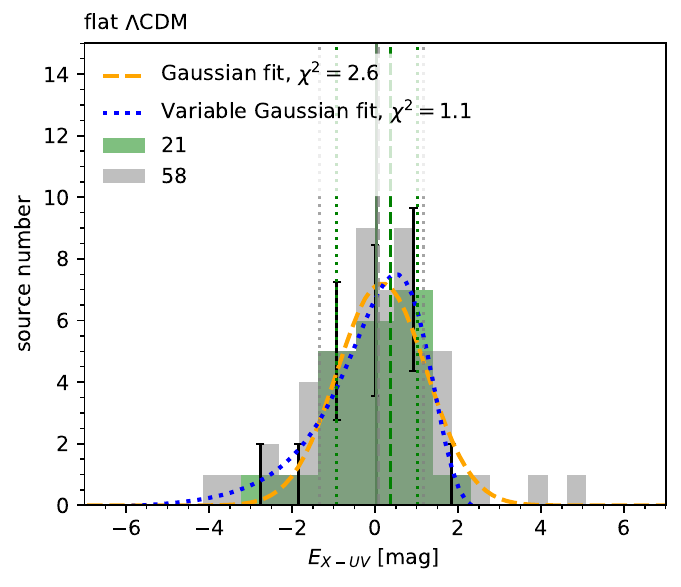}
    \caption{Unnormalized distributions of $\Delta \log{D_{\rm L}}=\log{D_{L,L_{X}-L_{UV}}}-\log{D_{L,R-L}}$ and $E_{X-UV}$ for the sample of 21 (58) sources in the flat $\Lambda$CDM model in green (grey). \textit{Left panel:} Distribution of the unnormalized luminosity-distance difference $\Delta \log{D_{\rm L}}=\log{D_{L,L_{X}-L_{UV}}}-\log{D_{L,R-L}}$ for the subsample of 21 sources. The solid vertical green (grey) line stands for the distribution mean, the dashed vertical green (grey) line represents the median, and the dotted vertical green (grey) lines stand for $16\%$ and $84\%$ percentiles. The best-fit Gaussian function is depicted by an orange dashed line, while the best-fit variable Gaussian function is represented by a dotted blue line.  \textit{Right panel:} The distribution of $E_{X-UV}$ (in magnitudes)  in green (grey) for the sample of 21 (58) sources. The vertical lines as well as the dotted blue and the dashed orange lines have the same meaning as in the left panel.}
    \label{fig_lum_distance_diff21_hist}
\end{figure*}

Luminosity distances for the 21 sources inferred using the two relations are shown in Fig.~\ref{fig_lum_distance}. We show the unnormalized distributions of the luminosity-distance difference in Fig.~\ref{fig_lum_distance_diff21_hist} (left panel). We list the basic statistical parameters of the unnormalized distribution in the last line of Table~\ref{tab_diff_dist}. The removal of extincted sources does affect the distribution by decreasing the mean value of the luminosity-distance difference. In addition, skewness becomes negative since the positive tail of the distribution disappears. Fisher's kurtosis also gets smaller because the obscured sources in the tails are removed. On the other hand, the median value increases which may be the effect of the small size of the subsample. We also fit the $\Delta \log{D_{\rm L}}$ and $E_{X-UV}$ distributions using the normal Gaussian and the variable Gaussian functions, see the left and the right panels in Fig.~\ref{fig_lum_distance_diff21_hist}, respectively, with the variable Gaussian fitting the distributions better according to the $\chi^2$ statistic. Actually, the distribution of $E_{X-UV}$ becomes even more asymmetric with $\sigma_{+}<\sigma_{-}$, see Table~\ref{tab_E_X_UV}, and the peak is positively shifted with $E_{X-UV}=0.50\pm 0.12$ mag, while the median is at $\sim 0.4$ mag. We thus see that the source pre-selection alleviates the problem of extinction by removing the outliers and reducing the tails. The change in the distribution characteristics between the samples of 58 and 21 sources indicates that extinction plays a role in increasing the distribution tails consisting of outliers and inducing a positive skewness. However, even after the application of hard X-ray and far-UV cuts, the positive shift of the distribution peak still indicates a contribution from extinction.       

\begin{figure}
    \centering
     \includegraphics[width=0.45\textwidth]{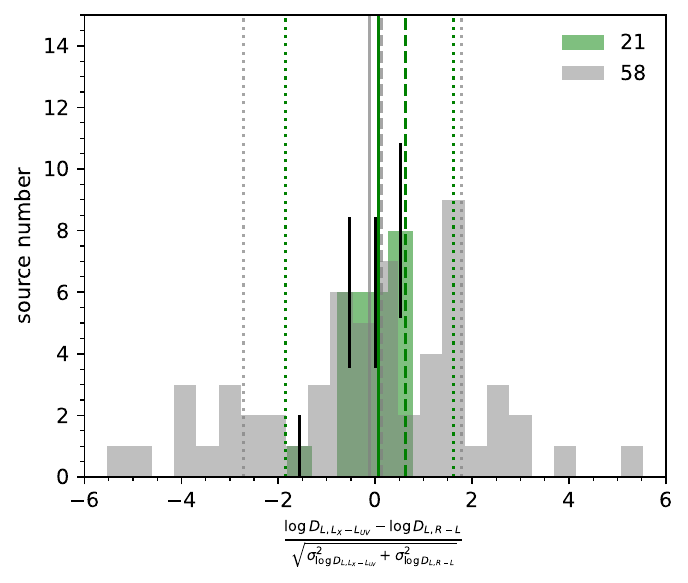}
    \caption{The distribution of $\Delta \log{D_{\rm L}}=\log{D_{L,L_{X}-L_{UV}}}-\log{D_{L,R-L}}$ normalized by the square root of the sum of the luminosity-distance uncertainties. The distributions are shown for the whole sample of 58 sources (grey) and the subsample of 21 sources (green). The vertical lines denote the same statistical properties as in Fig.~\ref{fig_lum_distance_diff21_hist}.}
    \label{fig_lum_distance_diff21_norm}
\end{figure}

The normalized distribution of luminosity-distance differences shown in Fig.~\ref{fig_lum_distance_diff21_norm} remains negatively skewed as for the bigger sample of 58 sources analyzed in \citet{Khadkaetal2023}. For the sample of 58 sources, the mean of the normalized distribution is negative ($-0.116$), which indicates that luminosity distances inferred from $L_{X}-L_{UV}$ relation are smaller than those inferred from the $R-L$ relation. This is consistent with the overall preference of larger $\om$ for the $L_{X}-L_{UV}$ relation constructed for 58 sources. The removal of extincted sources and the final subsample of 21 sources exhibit the normalized distribution with a positive mean value (+0.075), hence this indicates an opposite trend. However, since the subsample of 21 sources is more than a factor of two smaller in size, the change in the sign of the mean value may be the result of limited statistics and it does not reflect the trend in the larger samples. However, as for the unnormalized distribution, the absolute value of the mean decreases, which implies that the removal of extincted sources is beneficial for decreasing the luminosity-distance difference between $L_{X}-L_{UV}$ and $R-L$ relations.  

\subsection{Relation to the extinction curve and $E_{B-V}$}

Galactic extinction is generally corrected since the $(B-V)$ colour excess is known for the whole range of Galactic coordinates \citep[see e.g.][]{1998ApJ...500..525S}. Our sample of 58 sources is in a region in the sky that is far from the Galactic plane, which has Galactic extinction of $E_{B-V}\sim 0.1$ mag, see Fig.~\ref{fig_quasars_dustmap}. However, there still may be uncertainty in the applied Galactic extinction correction at the level of $E_{B-V}\sim 0.001$ mag. For instance, for NGC 4151 the colour index $E_{B-V}\sim 0.024$ mag according to \citet{2011ApJ...737..103S}, while it was measured to be $E_{B-V}\sim 0.027$ mag according to \citet{1998ApJ...500..525S}; see NED\footnote{\url{https://ned.ipac.caltech.edu/}} for the extinction values and their scatter for the other sources.

The more uncertain extinction contribution is due to the QSO optical, UV, and X-ray continuum emissions being affected by intrinsic extinction. For the Sloan Digital Sky Survey data (4576 QSOs), \citet{2003AJ....126.1131R} constructed composite spectra, which they categorized into six composite spectral classes (Composite 1 -- 6). Composite 1 corresponds to sources with intrinsically blue (or optically flat) power-law continua, while Composite 5 corresponds to reddened sources with optically steep power-law continua. Most of the quasars belong to rather bluer composite classes with colour excess $E_{B-V}<0.04$ mag, while only $\sim 10\%$ are reddened.

Based on the composite spectra constructed by \citet{2003AJ....126.1131R}, \citet{2004MNRAS.348L..54C} derived the extinction curve of QSOs, which is generally similar to the Small Magellanic Cloud extinction curve \citep{1984A&A...132..389P}, e.g. lacking the absorption feature at 2200\,\AA\, except for the shortest UV wavelengths. The QSO extinction curve can generally be attributed to a circumnuclear dusty shell composed of amorphous carbon grains that lack both silicate and graphite grains based on the corresponding missing spectral features.

In the following, we use the simplified analytical extinction curve of \citet{2004MNRAS.348L..54C},
\begin{equation}
    \frac{A_{\lambda}}{E_{B-V}}=-1.36+13\log{(1/\lambda\, [{\rm \mu m}])}\,,
    \label{eq_extinction_curve}
\end{equation}
and extrapolate it to the UV/X-ray range (2500\,\AA--2\,keV). Here $A_{\lambda}$ is an extinction correction at wavelength $\lambda$, and $E_{B-V}$ is the colour excess measured between B and V bands which are customarily used as the extinction measure in the optical/UV band. Subsequently, from the inferred UV/X-ray colour excess $E_{X-UV}=A_{X}-A_{UV}\sim 0.03-0.28$ mag (median and peak values, see Table~\ref{tab_E_X_UV}), we can use the relation $E_{X-UV}=13 E_{B-V}\log{(\lambda_{UV}/\lambda_{X})}$ derived from Eq.~\eqref{eq_extinction_curve} to obtain the colour excess of $E_{B-V}\sim 0.001-0.01$ mag. This can be interpreted to be predominantly the intrinsic colour excess expected for the majority of type I QSOs \citep{2003AJ....126.1131R}. This extinction originates in the circumnuclear medium, e.g. in an obscuring torus, a warped disk-like structure, or an outflowing clumpy wind within $\sim 1$ pc from the supermassive black hole \citep{2002ApJ...567L.107E,2017MNRAS.468.4944G,2018MNRAS.478.1660G,2023MNRAS.519.4082G}, and the host galaxy interstellar medium, also see \citet{2023MNRAS.522.2869S} and \citet{2023EPJD...77...56C} for discussions. Dust can also be present on the scales of a few thousand gravitational radii, i.e. on subparsec scales, within the BLR clouds \citep{2023arXiv231005089P}. Dusty structures can be located close to the SMBH on subparsec scales, especially for lower-luminosity sources. Considering the source at the low-luminosity end, see Table~\ref{tab_sample_description}, we have the UV luminosity of $\nu_{UV}L_{UV}\sim 1.2\times 10^{43}\,{\rm erg\,s^{-1}}$, which implies the sublimation radius of $r_{\rm sub}\sim 0.04(T_{\rm sub}/1500\,{\rm K})^{-2.8}(\nu_{UV}L_{UV}/10^{43}{\rm erg\,s^{-1}})^{1/2}\,{\rm pc}$ in the optically-thin limit \citep{1987ApJ...320..537B,2014A&A...565A..17Z}. Actually, in the low-luminosity limit of the Galactic center, compact dusty objects were detected on the scale of $\sim 1$ milliparsec \citep{2012Natur.481...51G,2021ApJ...923...69P}, hence understanding the 3D distribution and geometry of dust in galactic nuclei as a function of their accretion rate is relevant for estimating the intrinsic extinction in different wavebands. 

Hence the discrepancy between luminosity distances inferred using $L_{X}-L_{UV}$ and $R-L$ relations is expected for any selection of QSOs. This is consistent with the finding that the application of extinction cuts based on the hard X-ray index and the far-UV slope \citep{Lussoetal2020} only slightly mitigates the extinction problem, as we showed in Subsection~\ref{subsection_subsample}, by eliminating the outliers forming the tails. However, other qualitative properties of the $E_{X-UV}$ distribution, mainly the peak shifted to positive values, persist even after applying the extinction cuts.

We note that including many lower luminosity QSOs in the sample likely enhances the extinction effect. \citet{weaver2022} in their study of 9242 QSOs, all located in SDSS Stripe 82, derived a median extinction $E_{B-V}$ of 0.1 for redshifts around 2 and higher extinction values for lower and higher redshifts (see their Fig.\ A1).

Getting an estimate of the total reddening in AGN is crucial. The typical method involves hydrogen line ratios along UV, optical, and NIR wavelengths such as Ly$\alpha$/H$\beta$, H$\alpha$/H$\beta$ and Pa$\beta$/H$\beta$ \citep{2006agna.book.....O, 2023MNRAS.519.4082G, Panda_Sniegowska_2022arXiv220610056P}. Other methods require simultaneous measurements in a broad wavelength range \citep{shuder1979, choloniewski, 2007MNRAS.380..669C}. Unfortunately, none of these methods can be implemented for our sample due to the lack of broadband measurements. 

\section{Conclusions}
\label{sec:conclusions}

We find that the extinction (scattering and absorption) of X-ray and UV photons from QSOs contributes to the discrepancy between luminosity distances inferred using $L_{X}-L_{UV}$ and $R-L$ relations. 

For the non-zero luminosity-distance difference, i.e $\Delta \log{D_{\rm L}}=\log{D_{L,L_{X}-L_{UV}}}-\log{D_{L,R-L}}$, the extinction term is equal to $(\Delta \log{D_{\rm L}})_{\rm ext}=(\tau_X-\tau_{UV})\log{e}/[2(1-\gamma')]$ where $\tau_{X}$ and $\tau_{UV}$ are optical depths in the X-ray and UV domains, respectively, and $\gamma'$ is the slope of the $L_{X}-L_{UV}$ relation. We found that the distributions of $\Delta \log{D_{\rm L}}$ are asymmetric and positively shifted for all the six cosmological models considered. We estimated an average X-ray/UV colour index of $\overline{E}_{X-UV}=0.03-0.28$ mag in our sample, based on all six distribution median and peak values. We have shown that this amount of extinction is mild and overall typical for the majority of type I QSOs since it is supposed to originate in the circumnuclear and interstellar media of host galaxies \citep{2004MNRAS.348L..54C}. The dust-related systematic problem does not seem to be completely removed by standard hard X-ray and far-UV extinction cuts, hence some caution is necessary when interpreting the results \citep{KhadkaRatra2021a, KhadkaRatra2022}. Therefore, using at least two complementary methods for larger samples in the future is recommended.

\begin{acknowledgments}
 This research was supported in part by Dr.\ Richard Jelsma (a Bellarmine University donor), Research Foundation for the SUNY, US DOE grant DE-SC0011840, by the Polish Funding Agency National Science Centre, project 2017/26/A/ST9/00756 (Maestro 9), by GAČR EXPRO grant 21-13491X, by Millenium Nucleus NCN$19\_058$ (TITANs), and by the Conselho Nacional de Desenvolvimento Científico e Tecnológico (CNPq) Fellowships (164753/2020-6 and 313497/2022-2). BC and MZ  acknowledge the Czech-Polish mobility program (M\v{S}MT 8J20PL037 and PPN/BCZ/2019/1/00069). This project has received funding from the European Research Council (ERC) under the European Union’s Horizon 2020 research and innovation programme (grant agreement No. [951549]). Part of the computation for this project was performed on the Beocat Research Cluster at Kansas State University.
\end{acknowledgments}

%






\appendix

\section{Derivation of $D_{L,R-L}$, $D_{L,L_{X}-L_{UV}}$, and $E_{X-UV}$ expressions in the presence of extinction}
\label{appendix: formulae}

\subsection{Derivation of $D_{L,R-L}$}

Using the $R-L$ relation in the form,

\begin{equation}
    \log{\left(\frac{\tau}{{\rm days}}\right)}=\beta+\gamma \log{\left(\frac{L_{3000}}{10^{\eta}\,{\rm erg\,s^{-1}}} \right)}\,
    \label{eq_R-L_app}
\end{equation}
where the monochromatic luminosity $L_{3000}$ can be expressed in the form of the UV flux density at 2500\,\AA\, as $L_{\rm 3000,int}=4\pi D_{\rm L}^2 F_{3000,\nu}\nu_{3000}=4\pi D_{\rm L}^2 F_{UV}(2500/3000)^{\alpha_{\nu}}\nu_{3000}$, the $R-L$-based luminosity distance can be evaluated as 
\begin{equation}
    \log{D_{L,R-L}}=\frac{1}{2\gamma}(\log{\tau}-\beta)-\frac{1}{2}[\log{(4\pi)}-\eta+15.036]-\frac{1}{2}\log{F_{UV}}\,,
    \label{eq_DL_R-L_app1}
\end{equation}
where we used $\alpha_{\nu}\sim -0.45$ for the mean QSO continuum slope \citep{VandenBerk_etal_2020}. To extract the extinction term that modifies $\log{D_{L,R-L}}$, we use the extinction law in the form $F_{UV}=F_{\rm UV, int}e^{-\tau_{UV}}$ where $F_{\rm UV, int}$ is the intrinsic QSO UV flux density. Eq.~\eqref{eq_DL_R-L_app1} can then be rewritten as
\begin{equation}
    \log{D_{L,R-L}}=\frac{1}{2\gamma}(\log{\tau}-\beta)-\frac{1}{2}[\log{(4\pi)}-\eta+15.036]-\frac{1}{2}\log{F_{\rm UV, int}}+\frac{\log{e}}{2}\tau_{UV}\,,
    \label{eq_DL_R-L_app2}
\end{equation}
where the non-zero UV optical depth clearly increases $R-L$ luminosity distance with the term $+\frac{\log{e}}{2}\tau_{UV}\sim +0.2 \tau_{UV}$.

\subsection{Derivation of $D_{L,L_{X}-L_{UV}}$}

Analogously, from the $L_{X}-L_{UV}$ power-law relation parameterized as
\begin{equation}
    \log{\left(\frac{L_{X}}{{\rm erg\,s^{-1}\,Hz^{-1}}}\right)}=\beta'+\gamma'\log{\left(\frac{L_{UV}}{10^{\eta'}\,{\rm erg\,s^{-1}Hz^{-1}}}\right)}\,,
    \label{eq_UV-Xray_app}
\end{equation}
we can derive the $L_{X}-L_{UV}$-based luminosity distance $D_{L,L_{X}-L_{UV}}$ that depends on X-ray and UV monochromatic flux densities at 2 keV and 2500\,\AA\, $F_{X}$ and $F_{UV}$, respectively. The relation is as follows,
\begin{equation}
    \log{D_{L,L_{X}-L_{UV}}}=\frac{\beta'-\gamma' \eta'}{2(1-\gamma')}-\frac{\log{(4\pi)}}{2}+\frac{\gamma' \log{F_{UV}}}{2(1-\gamma')}-\frac{\log{F_{X}}}{2(1-\gamma')}\,.
    \label{eq_DL_UV-X-ray_app1}
\end{equation}
Using the extinction laws in the UV and the X-ray domains, $F_{UV}=F_{\rm UV, int}e^{-\tau_{UV}}$ and $F_{X}=F_{\rm X,int}e^{-\tau_{X}}$, respectively, we can express Eq.~\eqref{eq_DL_UV-X-ray_app1} as
\begin{equation}
    \log{D_{L,L_{X}-L_{UV}}}=\frac{\beta'-\gamma' \eta'}{2(1-\gamma')}-\frac{\log{(4\pi)}}{2}+\frac{\gamma' (\log{F_{\rm UV, int}}-\tau_{UV}\log{e})}{2(1-\gamma')}-\frac{\log{F_{\rm X,int}}-\tau_{X}\log{e}}{2(1-\gamma')}\,.
    \label{eq_DL_UV-X-ray_app2}
\end{equation}
By separating intrinsic flux density terms from the extinction terms, we obtain
\begin{equation}
    \log{D_{L,L_{X}-L_{UV}}}=\frac{\beta'-\gamma' \eta'}{2(1-\gamma')}-\frac{\log{(4\pi)}}{2}+\frac{\gamma' \log{F_{\rm UV, int}}-\log{F_{\rm X,int}}}{2(1-\gamma')}+\frac{\tau_{X}\log{e}}{2(1-\gamma')}-\frac{\tau_{UV}\gamma'\log{e}}{2(1-\gamma')}\,.
    \label{eq_DL_UV-X-ray_app3}
\end{equation}
hence for $\gamma'\sim 0.6$ extinction modifies $\log{D_{L,L_{X}-L_{UV}}}$ by $\sim +0.54\tau_{X}-0.33\tau_{UV}$, i.e. $L_{X}-L_{UV}$-based luminosity distance depends more strongly on the UV optical depth (it is decreased) than the $R-L$-based luminosity distance. In addition, it is increased by the non-zero X-ray optical depth, which is not present in the $R-L$-based luminosity distance relation. Also, it is essential to note that the extinction effect for the $L_{X}-L_{UV}$ relation depends on its slope $\gamma'$, hence this leads to a circularity problem in evaluating the extinction terms.

\subsection{Derivation of $E_{X-UV}$}

The colour index between X-ray and UV domains is defined as $E_{X-UV}\equiv A_{X}-A_{UV}=1.086(\tau_{X}-\tau_{UV})$. To obtain $E_{X-UV}$ we first calculate the difference $\Delta \log{D_{\rm L}}=\log{D_{L,L_{X}-L_{UV}}}-\log{D_{L,R-L}}=\log{(D_{L,L_{X}-L_{UV}}/D_{L,R-L})}$ using Eq.~\eqref{eq_DL_UV-X-ray_app3} and Eq.~\eqref{eq_DL_R-L_app2},
\begin{equation}
   \Delta \log{D_{\rm L}}=\underbrace{\overbrace{\frac{\beta'-\gamma' \eta'}{2(1-\gamma')}+\frac{\beta-\log{\tau}}{2\gamma}-\frac{\eta}{2}+7.518}^{\delta}+\frac{\log{F_{\rm UV, int}}-\log{F_{\rm X,int}}}{2(1-\gamma')}}_{=0\,\text{for intrinsic quasar emission}}+\underbrace{\frac{(\tau_{X}-\tau_{UV})\log{e}}{2(1-\gamma')}}_{\text{extinction contribution}}\,,
    \label{eq_DL_diff_factors_app}
\end{equation}
where the term $\delta+(\log{F_{\rm UV, int}}-\log{F_{\rm X,int}})/[2(1-\gamma')]$ is assumed to sum to zero for the intrinsic quasar emission, and hence the luminosity-distance difference for any source is zero without extinction. The non-zero difference is thus related to the extinction term, from which the optical-depth difference is $\tau_{X}-\tau_{UV}=2(1-\gamma')\Delta \log{D_{\rm L}}/\log{e}$. 

Finally, the colour index $E_{X-UV}$ can be expressed just as a function of the luminosity-distance difference and the slope $\gamma'$ of the $L_{X}-L_{UV}$ relation,
\begin{equation}
    E_{X-UV}=(2.172/\log{e})(1-\gamma')\Delta \log{D_{\rm L}}\simeq 5.001(1-\gamma')\Delta \log{D_{\rm L}}.
    \label{eq_colour_index_app}
\end{equation}

\section{$\delta$ versus $(\log{F_{UV}}-\log{F_{X}})/[2(1-\gamma')]$ distributions}
\label{appendix: delta-flux}

\begin{figure}
    \centering
    \includegraphics[width=0.48\textwidth]{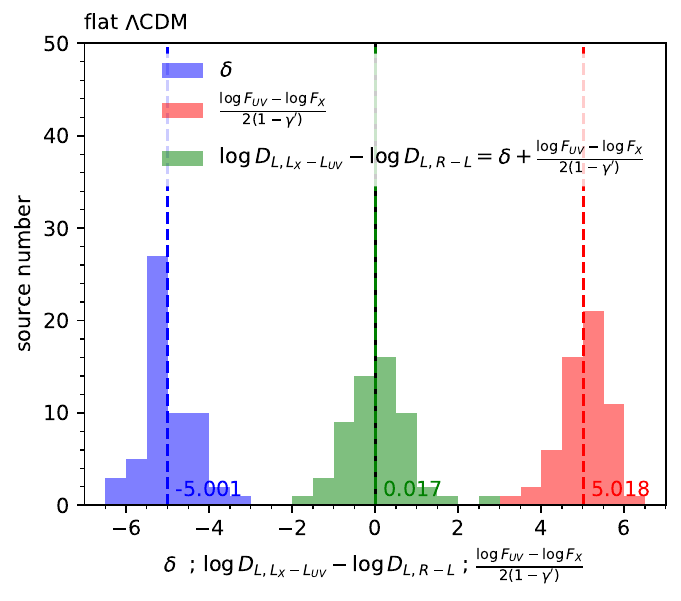}   
     \includegraphics[width=0.48\textwidth]{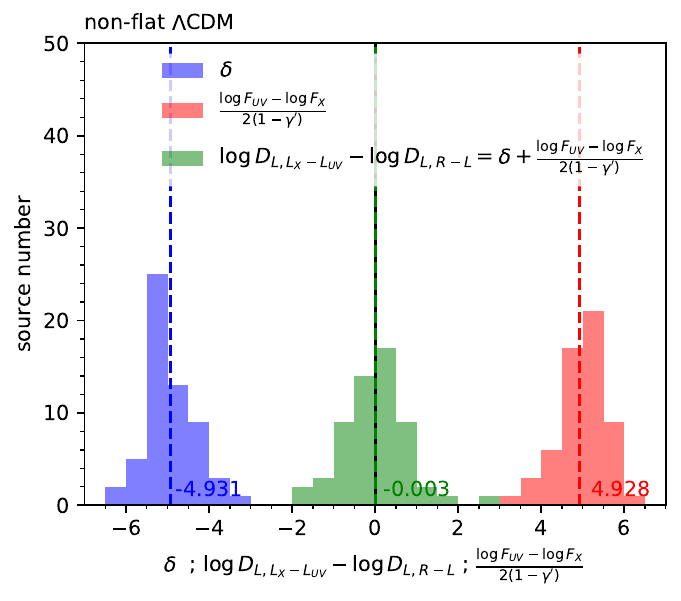}  
    \caption{Comparison of distributions of the factor $\delta$ (blue histogram; see Eq.~\eqref{eq_DL_diff_factors_app}), the factor $(\log{F_{UV}}-\log{F_{X}})/[2(1-\gamma')]$ (red histogram), and $ \log{D_{L,L_{X}-L_{UV}}}-\log{D_{L,R-L}}=\delta+(\log{F_{UV}}-\log{F_{X}})/[2(1-\gamma')]$ (green histogram) for 58 sources for flat and non-flat $\Lambda$CDM cosmological models in the left and the right panels, respectively. The black solid vertical line stands for zero, while the coloured dashed vertical lines (blue, green, and red) represent the means of the corresponding distributions.}
    \label{fig_delta_flux}
\end{figure}

To illustrate Eq.~\eqref{eq_DL_diff_factors_app}, i.e. $\Delta \log{D_{\rm L}}=\delta+(\log{F_{UV}}-\log{F_{X}})/[2(1-\gamma')]$, we compare distributions of $\delta$, $(\log{F_{UV}}-\log{F_{X}})/[2(1-\gamma')]$, and their sum $\Delta \log{D_{\rm L}}$ in Fig.~\ref{fig_delta_flux} for 58 sources and flat and non-flat $\Lambda$CDM models (distributions are similar for the other cosmological models). The mean value of $\delta$ is always negative, while the mean value of $(\log{F_{UV}}-\log{F_{X}})/[2(1-\gamma')]$ is positive and of a comparable magnitude. However, their sum, which is equivalent to the difference of luminosity-distance logarithms, is characterized by a distribution with positive skewness and non-zero mean in all cases (see Fig.~\ref{fig_delta_flux} and Table~\ref{tab_diff_dist} for details). This behaviour can qualitatively be attributed to the extinction term in Eq.~\eqref{eq_DL_diff_factors_app} proportional to $\tau_{X}-\tau_{UV}$, from which the colour index $E_{X-UV}$ can be quantified using Eq.~\eqref{eq_colour_index_app}. If $\tau_{X}>\tau_{UV}$, then the extinction term is positive and the distribution of $\Delta \log{D_{\rm L}}$ is shifted to positive values. The positive skewness is mainly given by the fact that for larger $\log{F_{UV}}-\log{F_{X}}$ the optical-depth difference tends to be greater since $\tau_{X}>\tau_{UV}$ for sources with a more absorbed X-ray emission.

\begin{figure}
    \centering
    \includegraphics[width=0.5\textwidth]{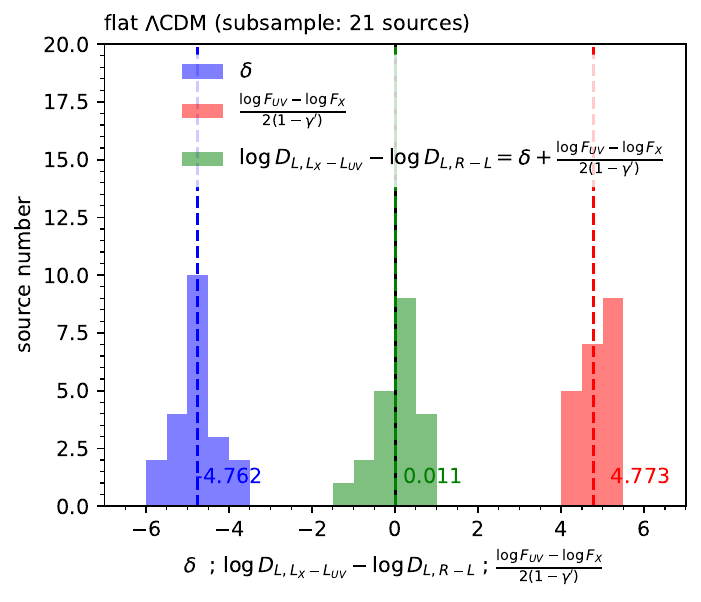}
    \caption{Comparison of distributions of the factor $\delta$ (blue histogram; see Eq.~\eqref{eq_DL_diff_factors_app}), the factor $(\log{F_{UV}}-\log{F_{X}})/[2(1-\gamma')]$ (red histogram), and $ \log{D_{L,L_{X}-L_{UV}}}-\log{D_{L,R-L}}=\delta+(\log{F_{UV}}-\log{F_{X}})/[2(1-\gamma')]$ (green histogram) for the subsample of 21 sources for the flat $\Lambda$CDM cosmological model. The black solid vertical line stands for zero, while the coloured dashed vertical lines (blue, green, and red) represent the means of the corresponding distributions.}
    \label{fig_delta_flux_21}
\end{figure}

For the sample of 21 sources that pass the reddening criteria of \citet{Lussoetal2020}, the mean value of  $\Delta \log{D_{\rm L}}=\delta+(\log{F_{UV}}-\log{F_{X}})/[2(1-\gamma')]$ decreases for the flat $\Lambda$CDM model in comparison with the full sample of 58 sources (0.011 versus 0.017), see Fig.~\ref{fig_delta_flux_21}. This implies that the reddening cuts are beneficial for decreasing the offset, however, the intrinsic extinction contribution seems to still be present.

\section{$E_{X-UV}$ distributions}
\label{appendix: E_X_UV}

\begin{figure}
    \centering
    \includegraphics[width=0.45\textwidth]{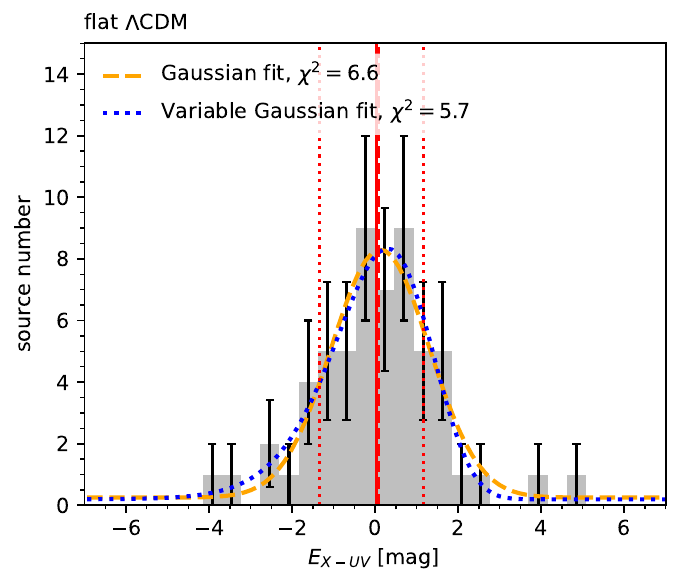}
    \includegraphics[width=0.45\textwidth]{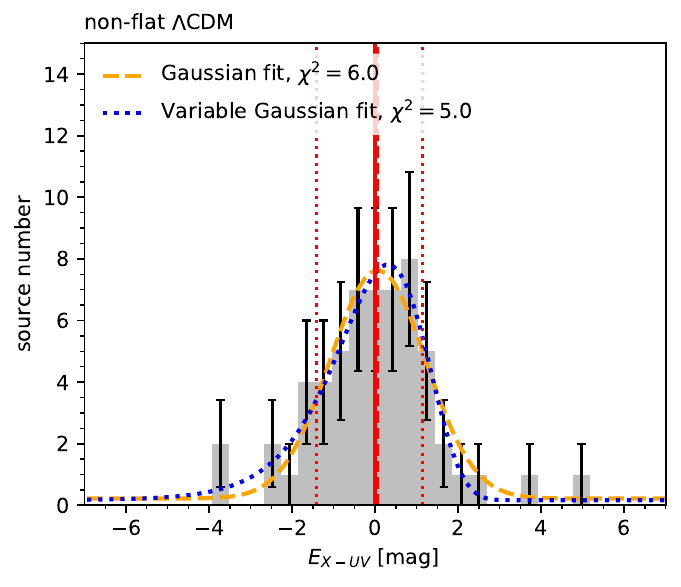}
     \includegraphics[width=0.45\textwidth]{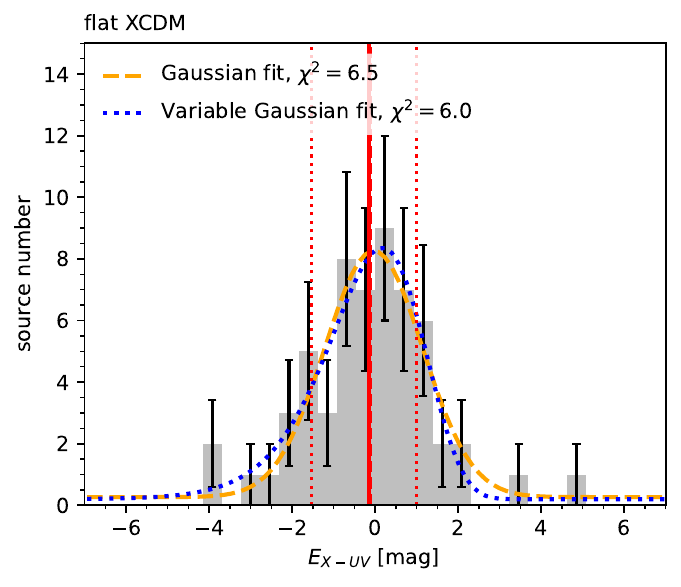}
    \includegraphics[width=0.45\textwidth]{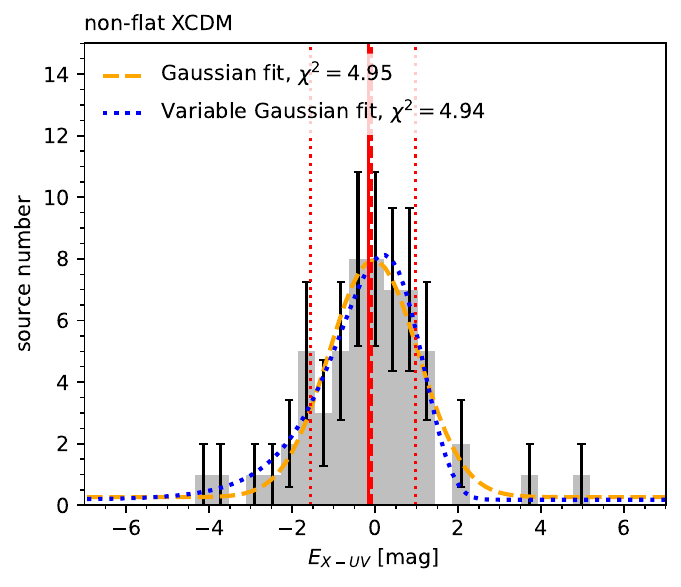}
    \includegraphics[width=0.45\textwidth]{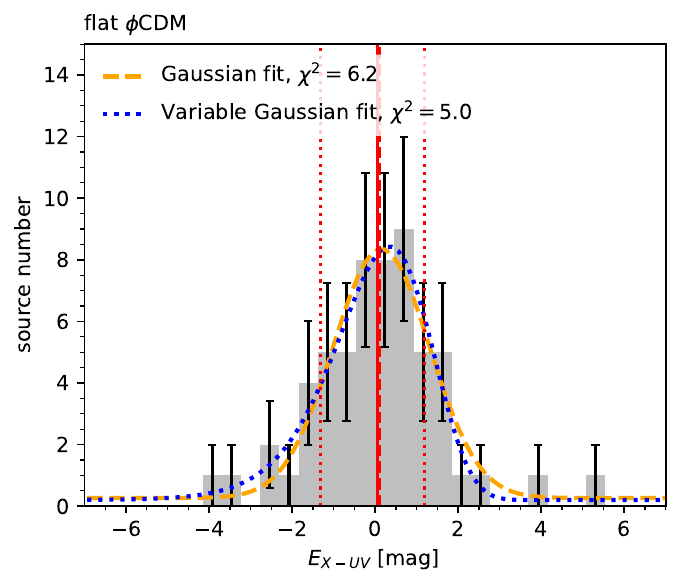}
    \includegraphics[width=0.45\textwidth]{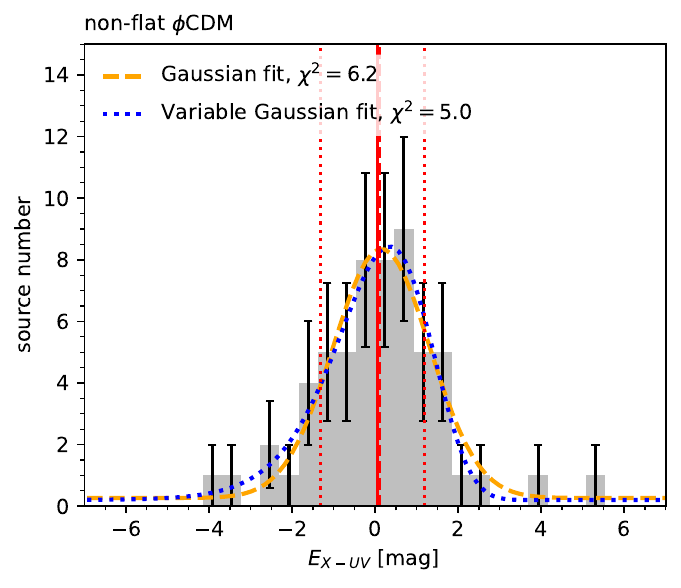} 
    \caption{Distributions of the X-ray/UV colour indices $E_{X-UV}$ for 58 sources for flat and non-flat $\Lambda$CDM, XCDM, and $\phi$CDM cosmological models (from top to bottom row). The X-ray/UV colour index $E_{X-UV}=5.001(1-\gamma')\Delta \log{D_{\rm L}}$ is calculated based on the luminosity distances for each source and the $L_{X}-L_{UV}$ relation slope for each cosmological model. Solid red vertical lines stand for $E_{X-UV}$ means, dashed red vertical lines are $E_{X-UV}$ medians, and dotted red vertical lines are the corresponding $16\%$ and $84\%$ percentiles. The bin width is set based on the Knuth binning algorithm and the histogram $y$ uncertainties for each bin are $\sigma_{y,i}=\sqrt{N_i}$ where $N_i$ is the number of points in each bin. The best-fit Gaussian function is represented by a dashed orange line, while the best-fit variable Gaussian function is depicted by a dotted blue line.}
    \label{fig_E_xuv}
\end{figure}

Using Eq.~\eqref{eq_Exuv_DeltaDL}, i.e. $E_{X-UV} \equiv A_{X}-A_{UV}\simeq 5.001(1-\gamma')\left\langle(\Delta \log{D_{\rm L}})_{\rm ext}\right\rangle$, we construct distributions of the colour index $E_{X-UV}$ from the luminosity-distance differences for each source. We show histograms of $E_{X-UV}$ for six cosmological models (flat and non-flat $\Lambda$CDM, XCDM, and $\phi$CDM models from the top to the bottom rows) in Fig.~\ref{fig_E_xuv}. All of the 58 sources have $E_{X-UV}$ in the interval ($-6$,6) and the histogram bin widths are determined based on the Knuth's rule. The red solid vertical line represents the distribution mean, the dashed vertical line stands for the median, and dotted red lines represent 16\%- and 84\%-percentiles. We also perform fits of Gaussian and variable Gaussian functions to all the distributions, which are shown by dashed orange and dotted blue lines, respectively.

\section{Inference of the reddening using the SDSS magnitudes}
\label{appendix:reddening}

\begin{figure}
    \centering
     \includegraphics[width=1.0\textwidth]{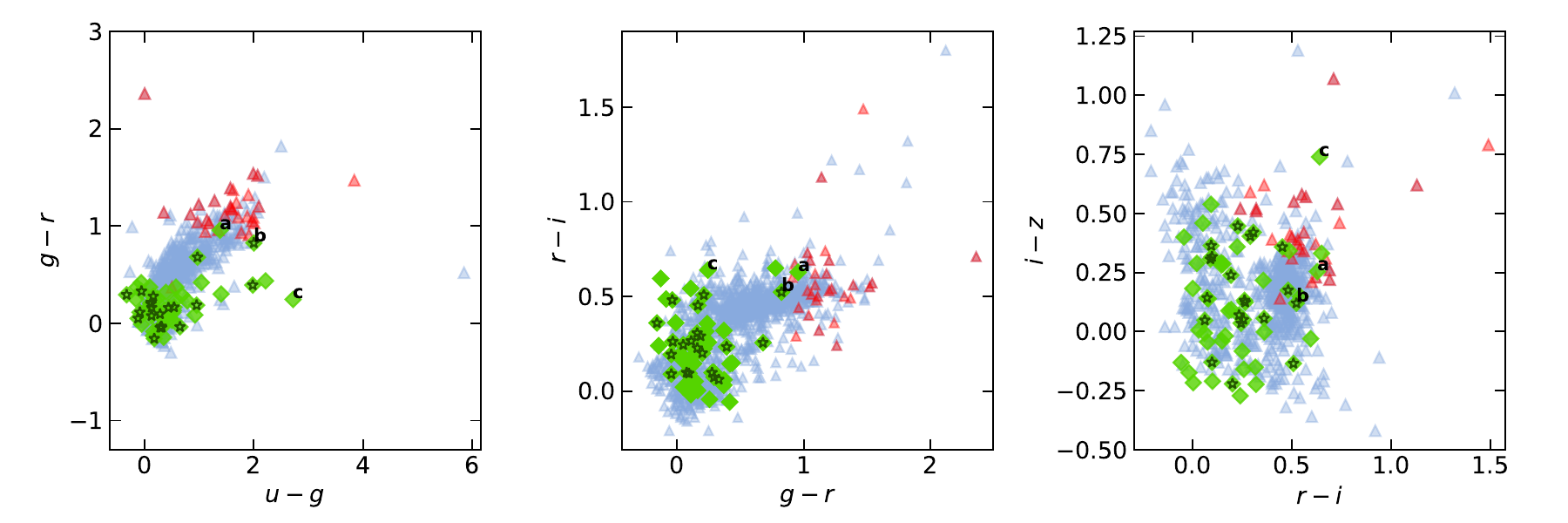}
    \caption{Colour-colour diagrams for blue and red quasar (triangle symbols) samples defined by \citet{glikman22}, and the \Mgii\ RM sample (green diamond symbols). Star symbols correspond to the 21 objects described in Section~\ref{sec:data}.  Lowercase letters identify the redder objects of the \Mgii\ RM sample: a) SDSS~J141110.95+524815.5
, b) SDSS~J142041.78+521701.6
, c) SDSS~J141645.58+534446.8. }
     \label{fig:sdss_colors}
\end{figure}

 An alternative possibility to infer the reddening effect in our sample involves using the \textit{ugriz} magnitudes from the SDSS database. To compare the behaviour of our sample, we have adopted the blue and red quasar samples defined by \citet{glikman22}. They considered red quasars as sources that have $E_{B-V}>0.25$. We have obtained the $u$ and $z$ magnitudes from the SDSS database, which are not reported by \cite{glikman22}. The SDSS magnitudes for the \Mgii\ RM sample were taken from \citet{shen_2019}. In all cases, the magnitudes were corrected for galactic extinction. Colour-colour plots are shown in Fig.~\ref{fig:sdss_colors}, where it is possible to observe the difference between the blue and the red samples. Most of the \Mgii\ RM sources are located at the left, bluer part of the distribution, which indicates that they have $E_{B-V}<0.25$. This result is consistent with the one found in Section~\ref{sec:results}. In Fig.~\ref{fig:sdss_colors} we also identify the 21 sources which satisfy the selection criteria of \citet{Lussoetal2020}, see Section~\ref{sec:data}; these exhibit the same behaviour as the full 58-source sample. Some of our sources are located in the red-quasar zone or show peculiar behaviour to the rest of the sample. We have identified three sources (SDSS~J141110.95+524815.5, SDSS~J142041.78+521701.6, and SDSS~J141645.58+534446.8) which are in the red zone or show different behaviour in at least two colour-colour diagrams. The consistent location of SDSS~J141110.95+524815.5 and SDSS~J142041.78+521701.6 in the red-quasar zone suggests that they have a high extinction. SDSS~J141645.58+534446.8 shows extreme behaviour, but it is not necessarily located in the red-quasar region. However, a visual inspection of the spectrum shows a flat continuum in the optical range, which suggests a high degree of reddening. One of these sources (SDSS~J142041.78+521701.6) belongs to the 21-source sample. Thus, the criteria based on hard X-ray index and far-UV slope criteria of \citet{Lussoetal2010} seem to be effective in cleaning the sample.

\begin{figure}
    \centering
     \includegraphics[width=0.65\textwidth]{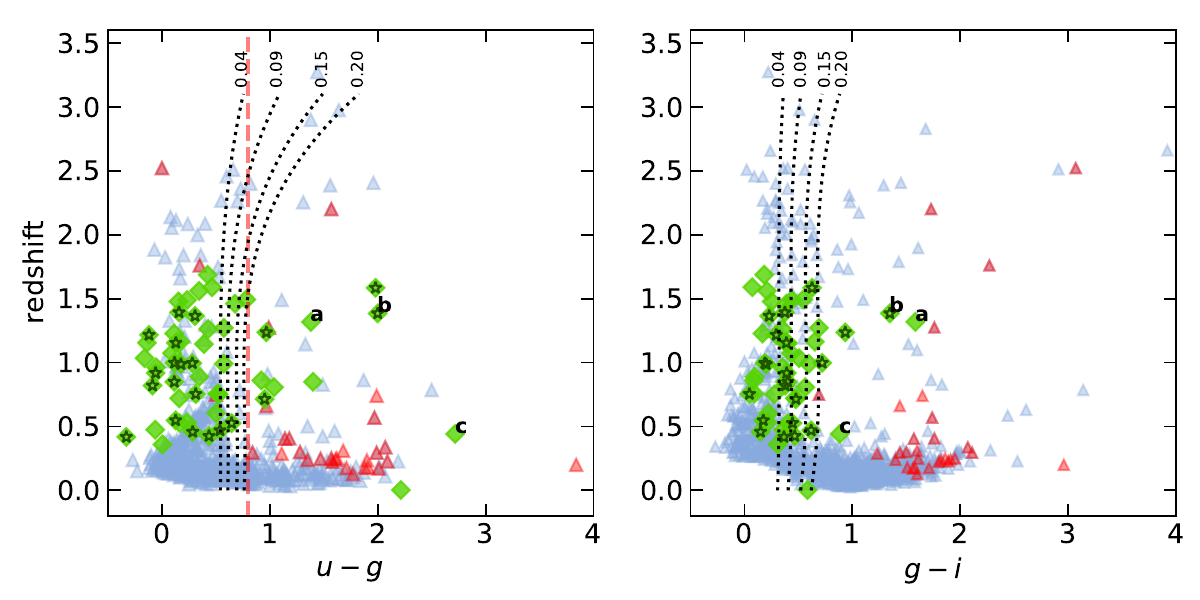}
    \caption{Redshift-colour diagrams. The dotted lines represent the expected change in relative colour as a function of redshift for $E_{B-V}=0.04, 0.09, 0.15$, and $0.20$ from left to right, respectively. To estimate $E_{B-V}$ change as a function of redshift, we used the extinction law of \citet{calzetti2000}. In the left panel, the vertical red-dashed line indicates the limit for red quasars according to \citet{2003AJ....126.1131R}  at $u-g=0.8$. Symbols have the same meaning as in Fig.~\ref{fig:sdss_colors}.}
     \label{fig:color-z}
\end{figure}

 Figure \ref{fig:color-z} shows the $u-g$ and $g-i$ distributions as a function of redshift. Using the \citet{calzetti2000} extinction law, we obtained the expected change in relative colour as a function of redshift for $E_{B-V}=0.04, 0.09, 0.15$, and $0.2$. We assumed a continuum slope of $\alpha=-1.56$ and $\alpha=-0.45$ following the composite spectrum of \citet{2001AJ....122..549V} for $u-g$ and $g-i$ colors, respectively. The behavior is the same as the one in Fig.~\ref{fig:sdss_colors}, red and blue quasar samples occupy different zones in the diagrams, and most of the \Mgii\ RM sources overlap with the blue quasar sample. The $E_{B-V}$ curves indicate that our sample has $E_{B-V}<0.2$. $E_{B-V}= 0.09$ is at the middle of the distribution, which supports the results found in Section~\ref{sec:results}.  In Fig.~\ref{fig:color-z} we also show the limit for red sources at $u-g=0.8$ according to \citet{2003AJ....126.1131R}, which coincides with the $E_{B-V}=0.2$ curve. Since most of the RM sample is to the left of this curve, this suggests that our sample is more consistent with a blue behaviour. The $E_{B-V}$ curves confirm the red colour of the three sources identified above. These objects should be excluded from future analyses. 

\end{document}